\documentclass[aps,physrev,reprint,noeprint]{revtex4-2}
\pdfoutput=1
 \usepackage{bm, braket,amsmath,mathtools,comment}
\usepackage[T1]{fontenc}
\usepackage{graphicx}
\usepackage{hyperref}

\begin{document}
\title{Matrix product state approach for a quantum system at finite temperatures using random phases and Trotter gates}
\author{Shimpei Goto}
\altaffiliation[Present address: ]{College of Liberal Arts and Sciences, Tokyo Medical and Dental University, Ichikawa, Chiba 272-0827, Japan}
\email[]{goto.las@tmd.ac.jp}
\author{Ryui Kaneko}
\email[]{rkaneko@phys.kindai.ac.jp}
\author{Ippei Danshita}
\email[]{danshita@phys.kindai.ac.jp}
\affiliation{Department of Physics, Kindai University, Higashi-Osaka city, Osaka 577-8502, Japan}
\date{\today}
\begin{abstract}
    We develop a numerical method based on matrix product states for simulating quantum many-body systems at finite temperatures without importance sampling and evaluate its performance in spin 1/2 systems.
    Our method is an extension of the random phase product state (RPPS) approach introduced recently [T.\ Iitaka, arXiv:2006.14459].
    We show that the original RPPS approach often gives unphysical values for thermodynamic quantities even in the Heisenberg chain.
    We find that by adding the operation of Trotter gates to the RPPS, the sampling efficiency of the approach significantly increases and its results are consistent with those of the purification approach.
    We also apply our method to a frustrated spin 1/2 system to exemplify that it can simulate a system in which the purification approach fails.
\end{abstract}
\maketitle
\section{Introduction\label{sec:introduction}}
Although it is unfeasible to simulate quantum many-body systems with classical computers in general, spatially one-dimensional (1D) systems are numerically tractable in spite of its strong quantum fluctuations.
This simulability relies on quite convenient tensor network representation, matrix product states (MPS)~\cite{schollwock_density-matrix_2011}.
The MPS have good natures for numerical simulations such as the existence of the canonical form and consequent numerical stability, and can faithfully represent a pure state with low entanglement.
Within the class of states with low entanglement, there exists some physically important subclasses.
For example, energetically low-lying states of 1D short-ranged systems do not possess extensive entanglement entropy~\cite{hastings_area_2007,hastings_entropy_2007,eisert_textitcolloquium_2010} and the famous density-matrix renormalization group (DMRG) algorithm~\cite{white_density_1992,white_density-matrix_1993,khemani_obtaining_2016} can reproduce these states.
Since the entanglement evolves continuously under unitary evolution, short-time dynamics starting from low entangled states is also simulatable with time-evolving block decimation algorithm or time-dependent DMRG~\cite{vidal_efficient_2003,vidal_efficient_2004,daley_time-dependent_2004,white_real-time_2004}.
Besides these pure states, mixed states at finite temperatures represented by the Gibbs states are tractable with classical computers by means of MPS~\cite{zwolak_mixed-state_2004,verstraete_matrix_2004,feiguin_finite-temperature_2005,white_minimally_2009,stoudenmire_minimally_2010}.

In order to represent the Gibbs states with MPS, there exists two typical approaches.
One is the purification approach in which the Hilbert space is enlarged~\cite{zwolak_mixed-state_2004,verstraete_matrix_2004,feiguin_finite-temperature_2005}.
A mixed state is represented as a pure state in the enlarged Hilbert space.
The other is the sampling approach in which pure states are randomly sampled from a mixed state~\cite{white_minimally_2009,stoudenmire_minimally_2010}.
The sampling approach does not require the enlarged Hilbert space and thus can be applied to more variety of systems than the purification approach.
For efficient sampling in the vast Hilbert space, the sampling approach is often combined with the importance sampling implemented by the Markov chain.
However, there is a drawback in the use of the Markov chain that it can bring the autocorrelation problem~\cite{lacki_dynamics_2015}, and thus some treatments to circumvent the problem are required in such approaches~\cite{white_minimally_2009,stoudenmire_minimally_2010,binder_symmetric_2017,chung_minimally_2019,chen_hybrid_2020,goto_quasiexact_2019,goto_minimally_2020}.

Recently, some sampling approaches without the Markov chain have been introduced~\cite{iwaki_thermal_2021,iitaka_random_2020}.
Without the Markov chain, the autocorrelation problem is absent and perfectly parallel computation is easily achievable.
Moreover, samples can be used to estimate the thermal averages of observables in a wide range of temperature in contrast to a method with the Markov chain, in which samples reproduce thermodynamics at a single temperature.

The absence of the Markov chain, however, also means the absence of the importance sampling.
It is questionable whether an MPS-based sampling approach without the importance sampling can efficiently represent statistics of the huge Hilbert space.
Although thermal pure quantum (TPQ) state approach~\cite{imada_quantum_1986,lloyd_pure_1988,jaklic_lanczos_1994,hams_fast_2000,sugiura_thermal_2012,sugiura_canonical_2013} can reproduce thermodynamics with a few samples, this success comes from the use  of typical states which MPS cannot efficiently represent because of volume-law entanglement~\cite{nakagawa_universality_2018}.

In this paper, we evaluate the performance of a simple sampling method without the importance sampling, namely, the random phase product state (RPPS) approach at first~\cite{iitaka_random_2020}.
The RPPS approach starts from a product state and its principle is quite a simple. This method is an adequate starting point to assess the sampling efficiency.
We define some indicators to quantify the sampling efficiency, and find that the efficiency of the RPPS approach is unsatisfactory even in the Heisenberg chain.
Moreover, the RPPS approach often gives unphysical negative values of the thermal entropy.
In order to resolve the sampling inefficiency and the unphysical result, we improve the RPPS approach by adding the operation of Trotter gates.
We dub the improved approach as random phase MPS with Trotter gates (RPMPS+T).
In the performance tests with the Heisenberg chain, we find that the sampling efficiency of the RPMPS+T approach increases by an order of magnitude and that its results are consistent with those of the purification approach.
Furthermore, we also perform a simulation in a frustrated spin 1/2 chain and the RPMPS+T approach succeeds in simulating its thermodynamics while the purification approach fails.

The rest of the paper is organized as follows: In Sec.~\ref{sec:methods}, we briefly review the RPPS approach and define indicators for the sampling efficiency.
On the basis of the indicators, we investigate the inefficiency of the RPPS approach and introduce the RPMPS+T approach to cure the inefficiency.
In Sec.~\ref{sec:results}, we present the performance tests of the RPMPS+T approach with the Heisenberg and frustrated zigzag chains.
In Sec.~\ref{sec:summaries}, we discuss how the RPMPS+T approach improves the sampling efficiency and the advantages of the RPMPS+T approach over existing approaches, and summarize the results.
 
\section{Entangled random phase matrix product state approach\label{sec:methods}}
\subsection{Model used for performance tests}
For the performance tests of the RPPS and our RPMPS+T approaches, we use a spin 1/2 antiferromagnetic \(J_1\)-\(J_2\) Heisenberg chain~\cite{okunishi_magnetic_2003,okunishi_calculation_2008,hotta_grand_2012}
\begin{align}
    \hat{H} = J_1 \sum^{L-1}_{i=1} \hat{\bm{S}}_i \cdot \hat{\bm{S}}_{i+1} + J_2 \sum^{L-2}_{i=1} \hat{\bm{S}}_i \cdot \hat{\bm{S}}_{i+2} - h\sum^L_{i=1} \hat{S}^z_i
\end{align}
with \(J_2 = 0\) (the Heisenberg chain) and \(J_2/J_1 = 1\) (the zigzag chain).
Here, \(J_1\) (\(J_2\)) is the strength of the spin-spin exchange interaction between neighboring (next-nearest neighboring) spins, \(h\) is the magnetic field, \(L\) is the number of lattice sites, and \(\hat{\bm{S}}_i = (\hat{S}^x_i, \hat{S}^y_i, \hat{S}^z_i)\) are the spin 1/2 operators at site \(i\).
\subsection{Random phase product state approach}
The RPPS approach introduced by \textcite{iitaka_random_2020} is a modification of the random phase state approach~\cite{iitaka_random_2004} to make it compatible with MPS.
In the random phase state approach, the trace of an operator \(\hat{O}\) is obtained as the sampling average of expectation values \(\braket{r|\hat{O}|r}\), where
\begin{align}
    \label{eq:random_phase_state}
    \ket{r} = \sum_\sigma \mathrm{e}^{i\theta_\sigma} \ket{\sigma}
\end{align}
is an unnormalized random phase state.
Here, \(\ket{\sigma}\) is an orthonormal basis, \(\theta_\sigma \) is a random number uniformly chosen from \((0, 2\pi]\), and the summation is taken over the basis.
Since the expectation value is expressed as 
\begin{align}
    \braket{r|\hat{O}|r} = \sum_\sigma \braket{\sigma |\hat{O}|\sigma} + \sum_{\sigma \neq \sigma^\prime} \mathrm{e}^{i(\theta_\sigma - \theta_{\sigma^\prime})}\braket{\sigma^\prime|\hat{O}|\sigma},
\end{align}
the phase factors become zero in the sampling average and thus 
\begin{align}
    \overline{\braket{r|\hat{O}|r}} = \mathrm{Tr} \hat{O}.
\end{align}
Here, the overline means the sampling average over random phase states \(\ket{r}\).
With this fact, one can obtain the thermal average of the operator \(\hat{O}\) in a system whose Hamiltonian is \(\hat{H}\) at inverse temperature \(\beta \) as
\begin{align}
    \begin{aligned}
        \ket{r_\beta} &= \mathrm{e}^{-\frac{\beta}{2}\hat{H}} \ket{r} \\
        \braket{\hat{O}}_\beta &= \frac{\mathrm{Tr}[\hat{O}\mathrm{e}^{-\beta \hat{H}}]}{A}\\
        &= \frac{\overline{\braket{r_\beta|\hat{O}|r_\beta}}}{\overline{\braket{r_\beta|r_\beta}}}.
    \end{aligned}
\end{align}
Here, \(A\) is the partition function \(A = \mathrm{Tr}\mathrm{e}^{-\beta \hat{H}}\).
Since the partition function \(A\) can be obtained as 
\begin{align}
A = \overline{\braket{r_\beta|r_\beta}},
\end{align}
any thermodynamic quantities can be obtained in this approach.
Depending on the orthonormal basis set, the random phase state approach can simulate both the canonical ensemble and the grand canonical ensemble.

Moreover, if one calculates the state \(\ket{r_\mathrm{\beta}}\) by the successive applications of an imaginary-time evolution operator with small imaginary-time step \(\mathrm{e}^{-\Delta \beta \hat{H}}\), states \(\ket{r_{m \Delta \beta}}\) for inverse temperature \(m \Delta \beta \leq \beta \) can be obtained during the imaginary-time evolution toward \(\beta \).
Here, \(m\) is the number of the operations.
Since the importance sampling is absent and thus the sampling does not target any temperature, these states can be utilized to estimate thermal expectation values and thermodynamic quantities at inverse temperatures up to \(\beta \).

On the other hand, MPS, 
\begin{align}
    \ket{\psi} = \sum_\sigma \bm{A}^{\sigma_1}_1 \bm{A}^{\sigma_2}_2 \ldots \bm{A}^{\sigma_L}_L \ket{\sigma},
\end{align}
is an efficient representation for low-entangled quantum states in spatially 1D systems.
Here, \(\sigma_i\) is the index of the local Hilbert space at site \(i\) and the orthonormal basis \(\ket{\sigma}\) is given by \(\bigotimes^L_{i=1} \ket{\sigma_i}\).
The efficiency of MPS representation can be inferred from the matrix dimensions of \(\bm{A}^{\sigma_i}_i\), which is often called as bond dimensions.
If the entanglement entropy with respect to the division of a system between sites \(i\) and \(i+1\) is \(S\), the bond dimension of \(\bm{A}^{\sigma_i}_i\) should be larger than \(\mathrm{e}^{S}\) for representing a state faithfully.
Therefore, MPS is an efficient representation of a quantum state in spatially 1D systems when the entanglement entropy of the state obeys the area-law scaling \(S \sim \mathrm{const}\).
If the entanglement entropy obeys the volume law scaling \(S \propto L\), the bond dimension should increase exponentially with system size \(L\) and thus numerical simulations are infeasible.

The random phase state in Eq.~\eqref{eq:random_phase_state} contains \textit{individual} random phase factors whose number is equivalent to the dimension of the Hilbert space, i.e., exponentially large.
Therefore, the MPS representation of the random phase state requires exponentially large bond dimensions in general, and the straight forward application of the random phase state approach to MPS representation is almost impossible.
In other words, the random phase state \(\ket{r}\) is a strongly entangled volume-law state.

In order to develop a variant of the random phase state approach in MPS representation, Iitaka has proposed the following RPPS,
\begin{align}
    \label{eq:RPPS}
    \ket{p} = \bigotimes^{L}_{i=1} \sum_{\sigma_i} \left(\mathrm{e}^{i \theta_{\sigma_i}}\ket{\sigma_i}\right).
\end{align}
Here, the summation is taken over the local Hilbert basis at site \(i\).
Since this is a product state, the bond dimensions of its MPS representation are only unity.
With the RPPS, one can obtain the thermal expectation value of an operator \(\hat{O}\) in the same way as the random phase state approach, i.e., 
\begin{align}
    \begin{aligned}
        \ket{p_\beta} &= \mathrm{e}^{-\frac{\beta}{2}\hat{H}} \ket{p} \\
        \braket{\hat{O}}_\beta &= \frac{\overline{\braket{p_\beta|\hat{O}|p_\beta}}}{\overline{\braket{p_\beta|p_\beta}}}.
    \end{aligned}
\end{align}
The RPPS in Eq.~\eqref{eq:RPPS} is the superposition of states in different Abelian symmetric sectors. Consequently, the RPPS approach simulates the grand canonical ensemble and the sample average of the squared norm \(\overline{\braket{p_\beta|p_\beta}}\) gives the grand canonical partition function \(\Xi \).

\subsection{Sampling inefficiency of random phase product state approach}
The RPPS approach will give correct thermal expectation values when sufficiently large samples are obtained.
However, compared to the minimally entangled typical thermal state (METTS) algorithm~\cite{white_minimally_2009,stoudenmire_minimally_2010} which also uses a sampling approach to calculate thermal expectation values, RPPS are selected uniformly and no importance sampling scheme is present.
The absence of the importance sampling could cause severe sampling inefficiency.

Let us investigate the sampling efficiency of the RPPS approach in the Heisenberg chain (\(J_2 = 0\)).
In order to quantify the efficiency of sampling, we first define the weight of each sample as follows: We denote an initial RPPS at \(l\)-th sample as \(\ket{p_l}\) and represent an expectation value of an operator \(\hat{O}\) from \(M\) samples as
\begin{align}
    \begin{aligned}
    \braket{\hat{O}}_{\beta, M} &= \frac{\frac{1}{M}\sum^M_{l=1}\braket{p_l|\mathrm{e}^{-\frac{\beta}{2} \hat{H}}\hat{O}\mathrm{e}^{-\frac{\beta}{2} \hat{H}}|p_l}}{\frac{1}{M}\sum^M_{k=1}\braket{p_k|\mathrm{e}^{-\beta \hat{H}}|p_k}} \\
    &= \sum^M_{l=1} \frac{\braket{p_l|\mathrm{e}^{-\frac{\beta}{2} \hat{H}}\hat{O}\mathrm{e}^{-\frac{\beta}{2} \hat{H}}|p_l}}{\braket{p_l|\mathrm{e}^{-\beta \hat{H}}|p_l}} \frac{\braket{p_l|\mathrm{e}^{-\beta \hat{H}}|p_l}}{\sum^M_{k=1}\braket{p_k|\mathrm{e}^{-\beta \hat{H}}|p_k}},
    \end{aligned}
\end{align}
and the weight of \(l\)-th sample is defined as
\begin{align}
    w_l = \frac{\braket{p_l|\mathrm{e}^{-\beta \hat{H}}|p_l}}{\sum^M_{k=1}\braket{p_k|\mathrm{e}^{-\beta \hat{H}}|p_k}}.
\end{align}
Since \(\braket{p_k|\mathrm{e}^{-\frac{\beta}{2} \hat{H}}\hat{O}\mathrm{e}^{-\frac{\beta}{2} \hat{H}}|p_k}/\braket{p_k|\mathrm{e}^{-\beta \hat{H}}|p_k}\) is an expectation value of an operator by a normalized wavefunction, the norm of this factor is bounded by the operator norm of \(\hat{O}\) which grows at most polynomially with system size in usual situations.
Meanwhile, the weight \(w_l\) can vary exponentially with inverse temperature \(\beta \) and system size: The distribution of the weights \(w_l\) is crucial for the efficiency of sampling.

\begin{figure}
    \includegraphics[width=\linewidth]{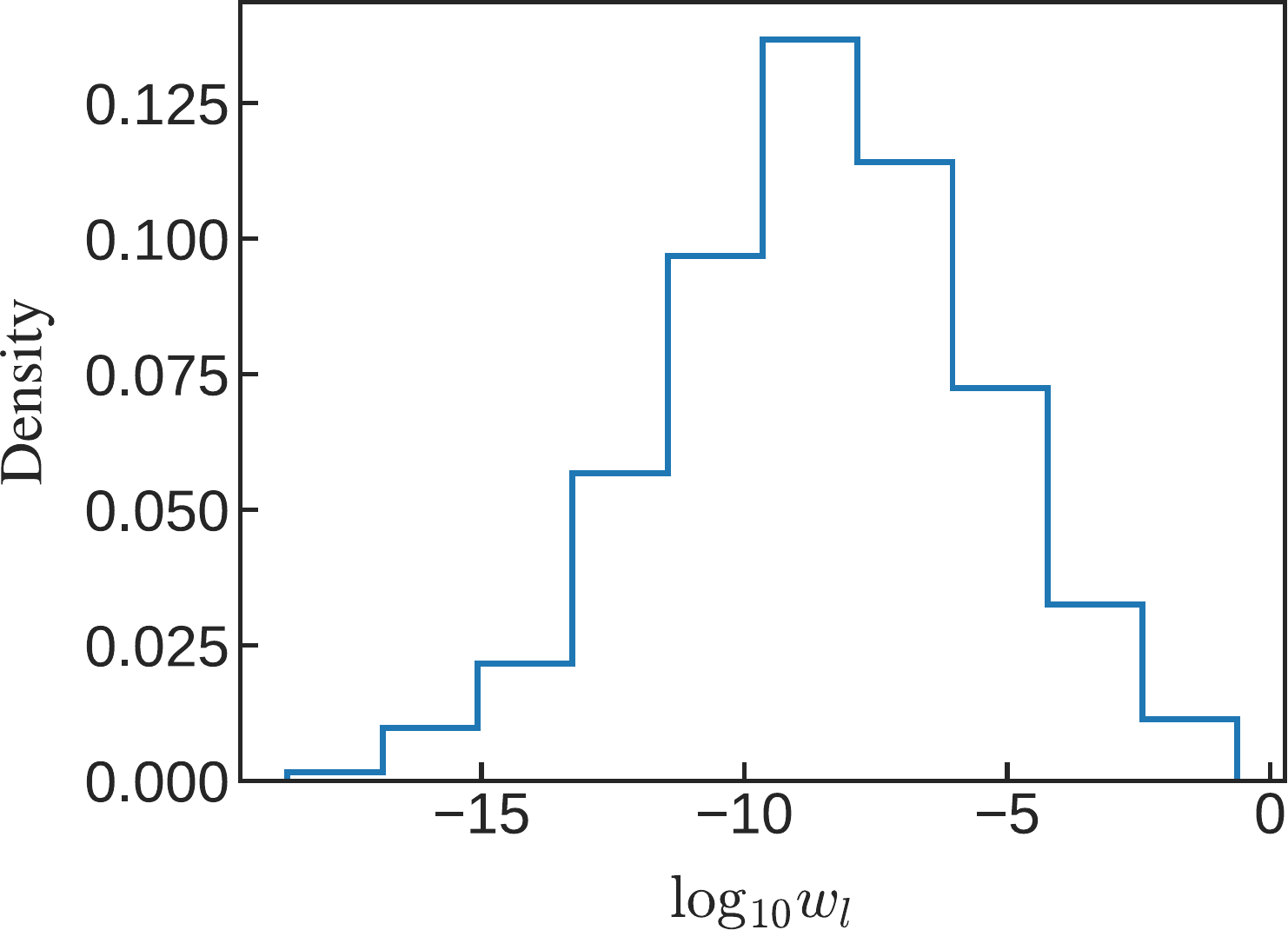}
    \caption{Histogram of sample weights \(w_l\) from the RPPS approach in the 100-site Heisenberg chain. The number of samples is 1024. The inverse temperature \(\beta \) is set to be 20 \(J^{-1}_1\).}\label{fig:hist_RPPS}
\end{figure}

Figure~\ref{fig:hist_RPPS} represents the histogram of 1024 sample weights \(w_l\) from the RPPS approach in the 100-site Heisenberg chain with \(h=0\) at \(\beta = 20J^{-1}_1\).
In the ideal sampling, every sample has almost the same weight and the peak of its histogram is located in \(1/1024 \approx 10^{-3}\).
On the contrary, the samples generated by the RPPS approach have exponentially varying weights as we have worried and the peak of the histogram is located in around \(10^{-9}\).
The sampling seems quite inefficient.

For quantitative discussions, we define the sampling efficiency from the generated \(M\) samples.
At first, weights \(w_l\) are reindexed in decreasing order and a monotonically increasing sequence \(f_n = \sum^n_{l=1}w_l\) is defined.
Next, one selects a positive real number \(q < 1\) and finds an integer \(n_q\) which is defined as the smallest number satisfying \(f_{n_q} > q\).
With \(M\), \(q\), and \(n_q\), we define the efficiency of a sampling \(\eta_q \) as
\begin{align}
    \eta_q = \frac{n_q}{q M}.
\end{align}
If \(q\) is close to 1, the efficiency \(\eta_q\) can be recognized as the fraction of samples which contributes to the \(100q\) percent of total weight.
In the case of the RPPS approach in the 100-site Heisenberg chain, the efficiency \(\eta_q\) with \(q=0.99\) is \(0.039\), i.e., only about four percent of samples contributes to the ninety-nine percent of total weight.
One can see the severe sampling inefficiency of the RPPS approach even in such a simple case.

Another efficiency indicator without the parameter \(q\) can be defined from the entropy of the weights,
\begin{align}
    S_w = - \sum_i w_i \ln w_i,
\end{align}
which quantifies the average information content of the sampling.
In the ideal sampling \(w_i = 1/M\), \(S_w = \ln M\), and thus \(M = \mathrm{e}^{S_w}\) holds.
Based on this relation, we convert the entropy of the weights to the effective sample number as
\begin{align}
    M_\mathrm{eff} = \mathrm{e}^{S_w},
\end{align}
and define the efficiency from the entropy \(\eta_S\) as
\begin{align}
    \eta_S = \frac{M_\mathrm{eff}}{M}.
\end{align}
In the case of the RPPS approach shown in Fig.~\ref{fig:hist_RPPS}, the effective sample number is \(12.9\) and the efficiency from the entropy is \(0.013\).
As well as in the value of \(\eta_q\), the sampling inefficiency of RPPS approach is clearly indicated in the small value of \(\eta_S\).

\subsection{Random phase matrix product states with Trotter gates approach}
In the previous subsection, we corroborated that the absence of the importance sampling in the RPPS approach causes severe sampling inefficiency as expected.
In order to seek a resolution of the problem of the sampling inefficiency, we pay attention to a famous and successful numerical method without the importance sampling to simulate quantum many-body systems at finite temperatures: The TPQ state approach~\cite{imada_quantum_1986,lloyd_pure_1988,jaklic_lanczos_1994,hams_fast_2000,sugiura_thermal_2012,sugiura_canonical_2013}.
In the TPQ state approach, the concept of typicality, expectation values of local operators given by typical states are very close to corresponding thermal expectation values, plays an important role, and sample dependencies are almost negligible if the size of Hilbert space is large enough.
However, a straight forward extension of the TPQ state approach to MPS is almost impossible: The reduced density matrices of typical states should be close to thermal density matrices, and thus the entanglement entropy should be close to the thermal entropy which obeys volume-law scaling.
Such highly entangled states cannot be efficiently described by MPS.

Nevertheless, the success of the TPQ state approach suggests a possible cure for the sampling inefficiency of the RPPS approach:
Using initially entangled states would improve the sampling efficiency because such an approach makes the RPPS approach slightly close to the TPQ state approach.
Following this expectation, we improve the RPPS approach in the following manner.
In order to make sampled initial states entangled, we operate a unitary transformation \(\hat{U}\) as
\begin{align}
    \label{eq:ERP}
    \ket{e} &= \hat{U} \ket{p},
\end{align}
and estimate the thermal expectation of an operator by
\begin{align}
    \begin{aligned}
        \ket{e_\beta} &= \mathrm{e}^{-\frac{\beta}{2}\hat{H}} \ket{e} \\
        \braket{\hat{O}}_\beta &= \frac{\overline{\braket{e_\beta|\hat{O}|e_\beta}}}{\overline{\braket{e_\beta|e_\beta}}}.
    \end{aligned}
\end{align}

Thanks to the similarity invariance of the trace operation, the unitary transformation \(\hat{U}\) in Eq.~\eqref{eq:ERP} does not affect sample averages.
On the contrary, the weight of \(l\)-th sample is given as
\begin{align}
\label{eq:weight_ERP}
w_l = \frac{\braket{p_l|\hat{U}^\dagger \mathrm{e}^{-\beta \hat{H}}\hat{U}|p_l}}{\sum^M_{k=1} \braket{p_k|\hat{U}^\dagger \mathrm{e}^{-\beta \hat{H}}\hat{U}|p_k}},
\end{align}
and thus each weight is affected by the unitary transformation unless the unitary transformation commutes with \(\mathrm{e}^{-\beta\hat{H}}\).

Likewise the purification method for the canonical ensemble~\cite{barthel_matrix_2016}, a random phase state in a certain Abelian symmetric sector can be constructed by enumerating possible quantum numbers at each bond of MPS and by assigning independent random phase factors \(\mathrm{e}^{i\theta}\) to symmetrically allowed elements.
A resulting state is an unnormalized randomly phased Dicke state~\cite{li_verification_2021}.
Starting from the randomly phased Dicke state instead of the RPPS, one can simulate the canonical ensemble within this framework.

Since the purpose of the unitary operation \(\hat{U}\) is to entangle an initial RPPS, the operation of some Trotter gates
\begin{align}
    \label{eq:TrotterU}
    \hat{U} = {\left[\mathrm{e}^{-i\tau \hat{H}^\prime_\mathrm{even}}\mathrm{e}^{-i\tau \hat{H}^\prime_\mathrm{odd}}\right]}^n
\end{align}
would be sufficient for this purpose.
Here, \(\tau \) is the time step for the Trotter gates, \(\hat{H}^\prime_\mathrm{odd}\) (\(\hat{H}^\prime_\mathrm{even}\)) is the odd (even) bond terms of the Hamiltonian for the Trotter gates \(\hat{H}^\prime \), and \(n\) is the number of the operations.
The Trotter Hamiltonian \(\hat{H}^\prime \) should be different from the system Hamiltonian \(\hat{H}\), so we adopt a spin 1/2 antiferromagnetic XXZ chain
\begin{align}
    \hat{H}^\prime = J \sum^{L-1}_{i=1} (\hat{S}^x_i \hat{S}^x_{i+1}+ \hat{S}^y_i \hat{S}^y_{i+1} + \Delta \hat{S}^z_i \hat{S}^z_{i+1})
\end{align}
with the anisotropy parameter \(\Delta \) for \(\hat{H}^\prime \).
This model is a simple 1D quantum spin chain with a controllable parameter and thus a suitable starting point for investigating the effects of the Trotter gates.
Here, \(J\) is the strength of the spin-spin exchange interaction between neighboring spins.
One can control the ``distance'' of \(\hat{H}^\prime \) and \(\hat{H}\) by the anisotropy parameter \(\Delta \).
Since the large anisotropy parameter makes \(\hat{H}^\prime \) quite different from \(\hat{H}\), it is naively expected to improve the sampling efficiency.
However, too large \(\Delta \) makes \(\hat{H}^\prime \) close to the classical Ising model and the improvement by quantum entanglement would be small.
Hence, the optimization of the parameter is required.
 
\section{Results of performance tests\label{sec:results}}
In this section, we perform the benchmark tests of RPMPS+T approaches.
For imaginary-time evolution in all MPS simulations, we use the second-order Suzuki-Trotter decomposition with imaginary time step \(0.025J_1^{-1}\) and set the truncation error to \(10^{-8}\).
The next-nearest-neighbor interaction is implemented by swap gates~\cite{stoudenmire_minimally_2010}.
The time step \(\tau \) for the Trotter gates \(\hat{U}\) in Eq.~\eqref{eq:TrotterU} is set to be \(0.5J^{-1}\). 
We perform all the simulations in this section on a single thread of Intel Xeon Gold 6238R processor.

\subsection{Heisenberg chain}
We first perform a benchmark test in the 100-site Heisenberg chain with \(h = 0\) and compare the sampling efficiency of the RPMPS+T approaches to that of the RPPS approach.

\begin{figure}
    \includegraphics[width=\linewidth]{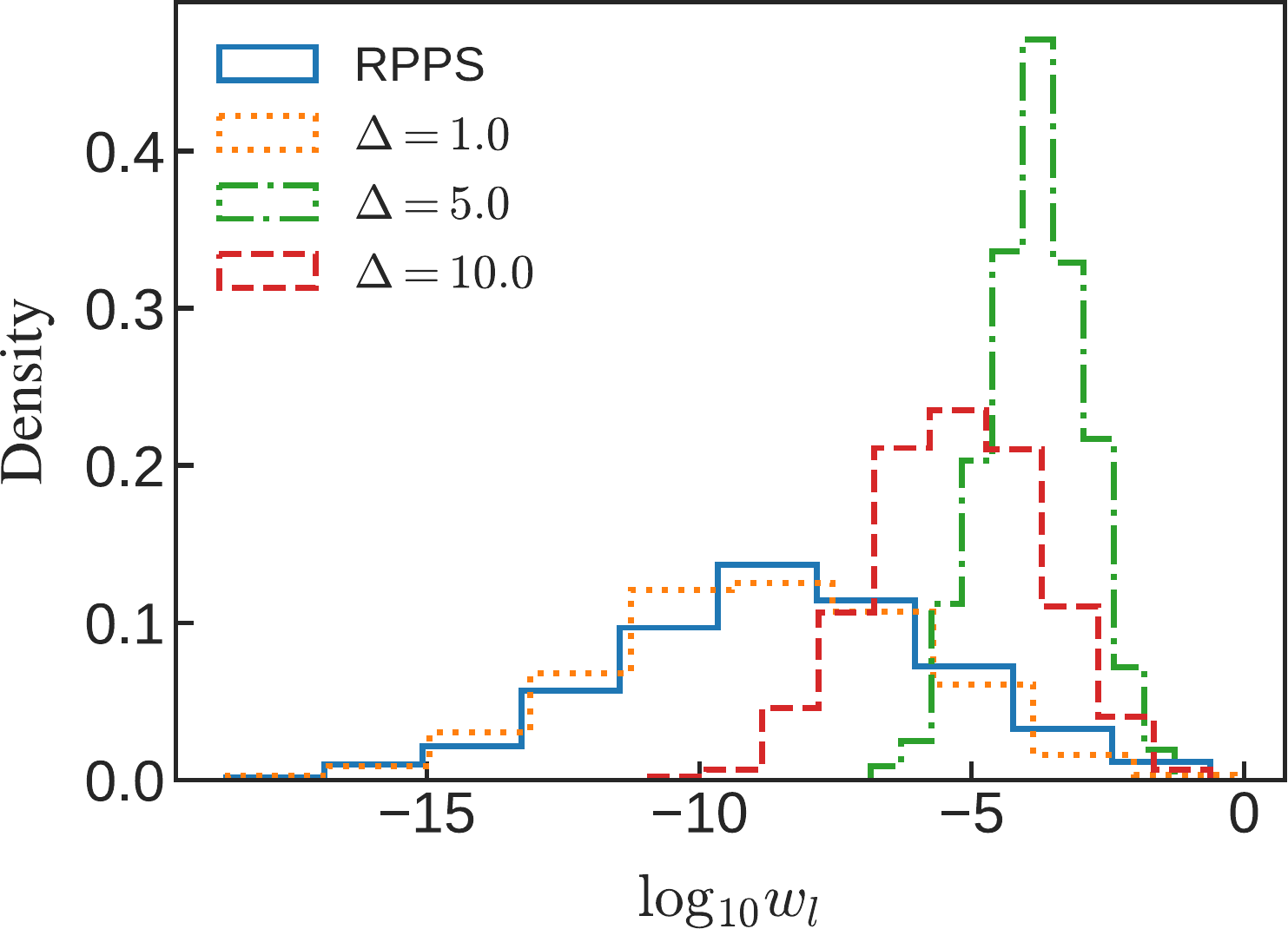}
    \caption{Histogram of sample weights \(w_l\) from the RPPS approach and the RPMPS+T approach in the 100-site Heisenberg chain. The number of samples is 1024. The inverse temperature \(\beta \) is set to be 20 \(J^{-1}_1\). The blue solid line corresponds to the RPPS approach. For the Trotter gates in the RPMPS+T approach, we use \(\Delta = 1.0\) (orange dotted line), 5.0 (green dashed-dotted line), and 10.0 (red dashed line). We apply the gates only once.\label{fig:hist_Jz}}
\end{figure}

Figure~\ref{fig:hist_Jz} represents the histograms of 1024 sample weights at inverse temperature \(\beta = 20 J^{-1}_1\) from the RPPS approach and the RPMPS+T approach with different anisotropy parameters.
With \(\Delta = 5.0\) (green dashed-dotted in Fig.~\ref{fig:hist_Jz}), the histogram becomes narrower than the RPPS case (blue solid line in Fig.~\ref{fig:hist_Jz}) and its peak is located around \(10^{-4}\).
The efficiency \(\eta_q\) with \(q=0.99\) increases up to 0.646 from 0.039 of the RPPS approach.
Another efficiency \(\eta_S\) also increases up to 0.218 from 0.013.
From these facts, one can say that the sampling efficiency is significantly improved.
The average computation time required to obtain one sample with the RPMPS+T approach up to \(\beta = 50.0 J^{-1}_1\) does not change so much from that of the RPPS approach: 69.6 seconds in the RPPS approach and 76.5 seconds in the RPMPS+T approach.
Therefore, the problem of the sampling inefficiency in the RPPS approach can be resolved by adding operation of proper Trotter gates to the RPPS approach.
As can be inferred from Eq.~\eqref{eq:weight_ERP}, the Trotter gate obtained from the original Hamiltonian (orange dashed lin in Fig.~\ref{fig:hist_Jz}) hardly affects the improvement of sampling efficiency.
Too large anisotropy parameter \(\Delta \) is not a good choice either as shown in \(\Delta = 10.0\) case in Fig.~\ref{fig:hist_Jz}.
We note that the sampling efficiencies from the Trotter gates with \((\Delta, n) = (7.0, 1)\) are \((\eta_{q=0.99}, \eta_S) = (0.628, 0.171)\) in the same system. Therefore, fine tunings of parameters are not necessarily required.

\begin{table}
    \caption{Indicators of the sampling efficiency \(\eta_q\) with \(q=0.99\) and \(\eta_S\) for different number of the unitary operations \(n\). For the unitary operation, we use the Trotter gates obtained from the antiferromagnetic XXZ chain with the anisotropic parameter \(\Delta = 5.0\). The averaged computation time for obtaining one sample \(t\) is also listed.  The one-sigma uncertainties of the indicators are estimated from the bootstrap analysis with 4000 resampled data.\label{table:efficiency}}
    \begin{ruledtabular}
        \begin{tabular}{cccc}
            \(n\) & \(\eta_q\) & \(\eta_S\) & \(t\) (s) \\
            \hline
            0 & 0.039\(\pm\)0.007 & 0.013\(\pm\)0.003 & 69.6\\
            1 & 0.646\(\pm\)0.015 & 0.218\(\pm\)0.022 & 76.5\\
            2 & 0.285\(\pm\)0.029 & 0.048\(\pm\)0.008 & 89.3\\
            3 & 0.229\(\pm\)0.044 & 0.026\(\pm\)0.013 & 199.1\\
            4 & 0.383\(\pm\)0.035 & 0.057\(\pm\)0.018 & 300.0\\
            5 & 0.635\(\pm\)0.035 & 0.149\(\pm\)0.072 & 481.1\\
            8 & 0.589\(\pm\)0.021 & 0.158\(\pm\)0.027 & 971.3\\
            10 & 0.631\(\pm\)0.031 & 0.113\(\pm\)0.031 & 2560.1\\
        \end{tabular}
    \end{ruledtabular}
\end{table}
With fixed anisotropy parameter \(\Delta=5.0\), we also investigate how the number of the unitary operation \(n\) affects the sampling efficiency.
Table~\ref{table:efficiency} summarizes the indicator of the sampling efficiency \(\eta_q\) with \(q=0.99\) and \(\eta_S\) for different \(n\).
The sampling efficiency is remarkably high at \(n=1\), drops at \(n=2\), and then increases gradually with \(n\) for \(3 \leq n \leq 5\).
The increased efficiency is comparable to that at \(n=1\).
In short, the sampling efficiency tends to be high with large \(n\) but comparable efficiency can be accomplished with the single operation with proper Trotter gates. 
Hereafter, we use \((\Delta, n) = (5.0, 1)\) for the Trotter gates.

\begin{figure}
    \includegraphics[width=\linewidth]{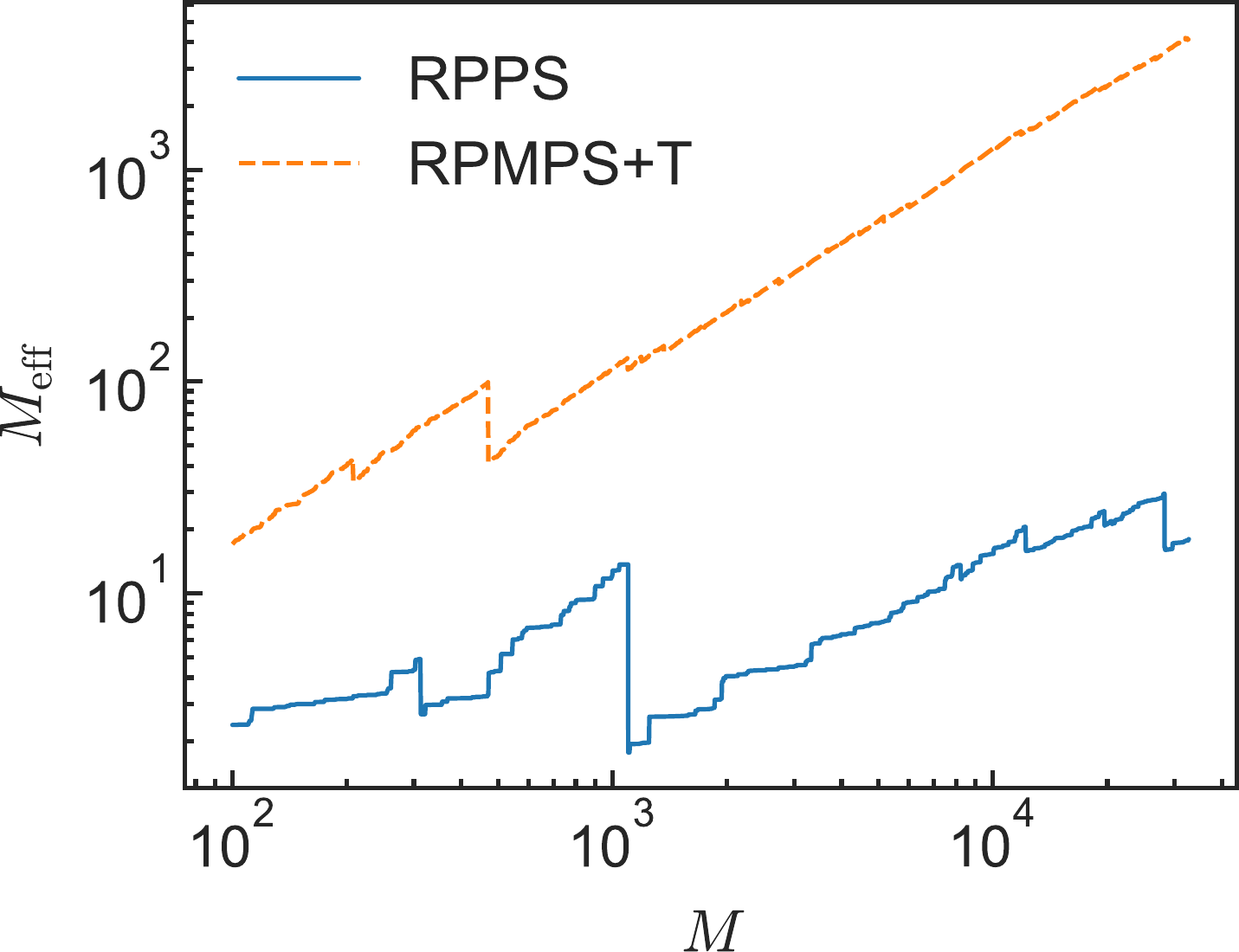}
    \caption{Effective sample size \(M_\mathrm{eff}\) as a function of the sample size \(M\) in the RPPS approach (blue solid line) and the RPMPS+T approach (orange dashed line) at \(\beta=20J^{-1}_1\) in the 100-site Heisenberg chain.\label{fig:Meff}}
\end{figure}

The difference between the RPPS approach and the RPMPS+T approach is quite visible in the sample size \(M\) dependence of the effective sample size \(M_\mathrm{eff}\) shown in Fig.~\ref{fig:Meff}.
The effective sample size increases almost linearly with the sample size except some sudden decreases coming from samples with extraordinary large weights.
Such samples with large weights effectively reduce the information content of former samples.
In the RPPS approach, we observe a sudden decrease around \(M \sim 10^3\), which is comparable to the effective sample size at that point.
On the contrary, the sudden decreases in the RPMPS+T approach destroy much less severely the information content of former samples.
Moreover, the increase of the effective sample size in the RPPS approach is quite slower than that in the RPMPS+T approach.

\begin{figure}
    \includegraphics[width=\linewidth]{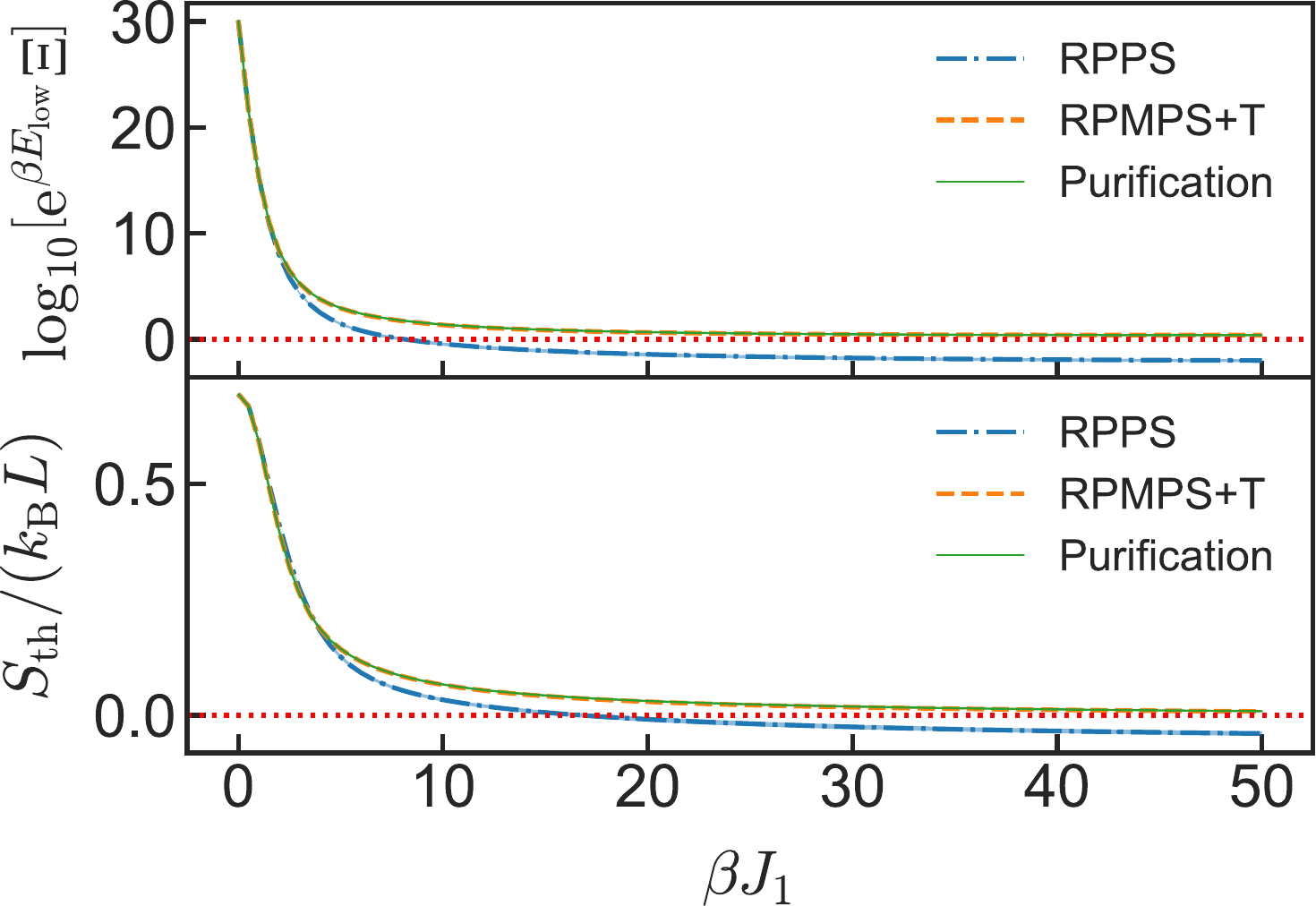}
    \caption{Inverse temperature dependence of the normalized partition function (the upper panel) and the thermal entropy (the lower panel) obtained by the RPPS approach (blue dashed-dotted line) and the RPMPS+T approach (orange dashed line) with 1024 samples in the 100-site Heisenberg chain. The values given by the purification approach with the same truncation error (green solid line) are also shown for reference. Shaded regions indicate one-sigma uncertainty. The one-sigma uncertainty of the thermal entropy is estimated from the jackknife analysis. The red dotted lines in both figures indicate the physically possible lowest values.~\label{fig:ZandS}}
\end{figure}

Next, we discuss how the efficiency affects thermodynamic quantities estimated from sampling.
The upper panel of Fig.~\ref{fig:ZandS} shows the partition functions obtained by the RPPS approach, the RPMPS+T approach, and the purification approach.
In order to avoid numerical overflow, the partition function is normalized as \(\mathrm{e}^{\beta E_\mathrm{low}}\Xi \), where \(E_\mathrm{low}\) is an approximated ground energy which is not lower than the true ground energy \(E_0\).
Here, we use \(\braket{\hat{H}}_\beta \) obtained by the purification approach at \(\beta = 50 J^{-1}_1\) for \(E_\mathrm{low}\).
The normalized partition function should be larger than 1:
\begin{align}
    \label{eq:ineq_partition}
    \mathrm{e}^{\beta E_\mathrm{low}}\Xi \geq \mathrm{e}^{\beta E_0}\Xi 
    = \mathrm{Tr}\left[\mathrm{e}^{-\beta(\hat{H} - E_0)} \right] 
    > 1.
\end{align}
Nevertheless, the normalized partition function obtained by the RPPS approach (blue dashed-dotted line in the upper panel of Fig.~\ref{fig:ZandS}) becomes less than 1.
The violation of the inequality~\eqref{eq:ineq_partition} means that the RPPS approach does not correctly capture the partition function due to the insufficient sampling.
It is worth noting that the error bars are no longer reliable when the sampling is inefficient as presented in the upper panel of Fig.~\ref{fig:ZandS}.
On the contrary, the normalized function obtained by the RPMPS+T approach fulfills the inequality~\eqref{eq:ineq_partition} and coincides with that obtained by the purification approach.

Thermodynamic quantities derived from the partition function would inherit its improper behavior.
The lower panel of Fig.~\ref{fig:ZandS} depicts the thermal entropy obtained by the RPPS approach, the RPMPS+T approach, and the purification approach.
The thermal entropy can be obtained as
\begin{align}
    S_\mathrm{th} = k_\mathrm{B} \left(\beta \braket{\hat{H}}_\beta + \ln Z\right).
\end{align}
Here, \(k_\mathrm{B}\) is the Boltzmann constant.
Although the thermal entropy should be nonnegative in the statistical mechanics, that obtained by the RPPS approach (blue dashed-dotted line in the lower panel of Fig.~\ref{fig:ZandS}) becomes negative even when the uncertainty is taken into account.
On the other hand, the RPMPS+T approach gives proper positive values of the thermal entropy (orange dashed line in the lower panel of Fig.~\ref{fig:ZandS}).

\begin{figure}
    \includegraphics[width=\linewidth]{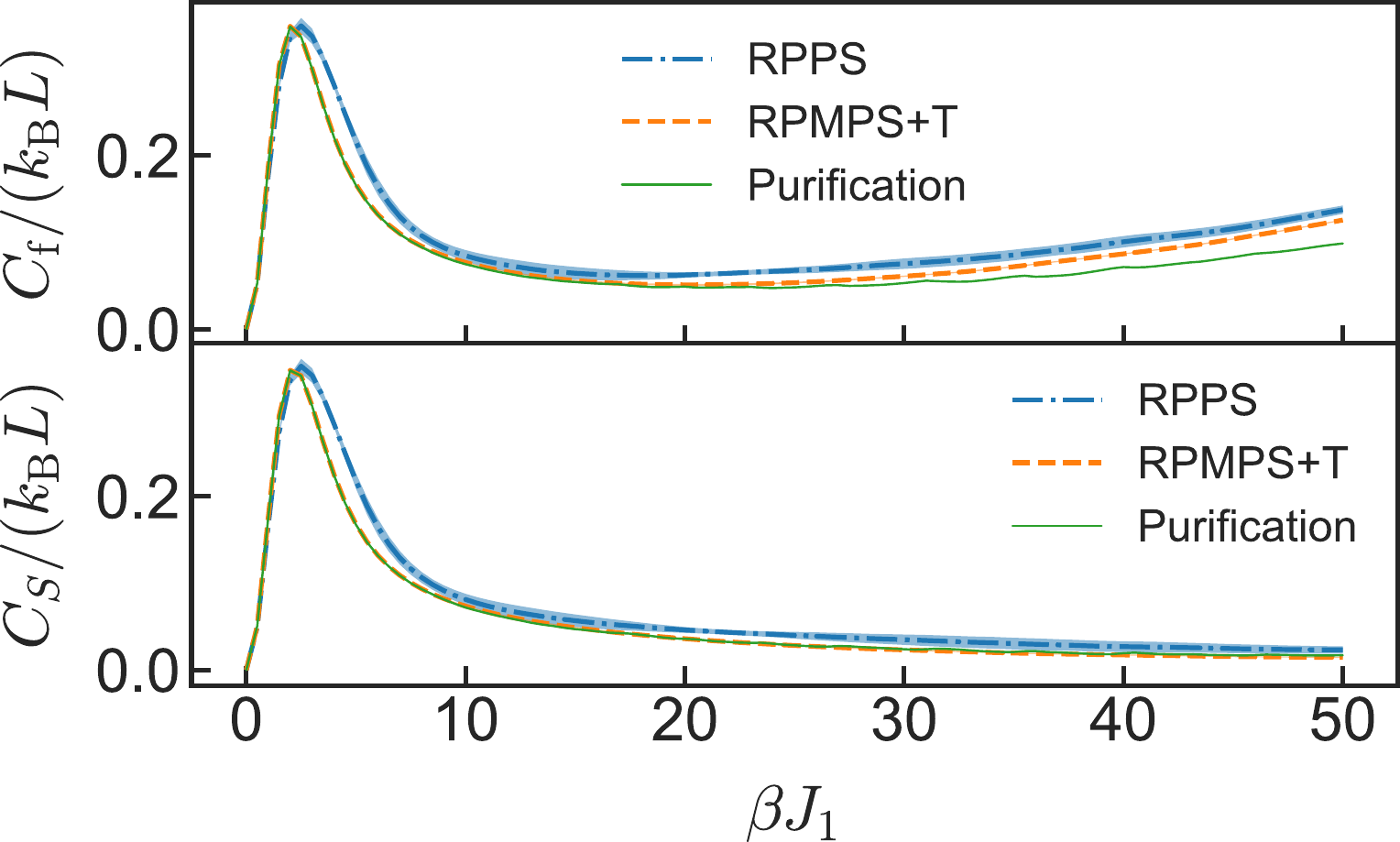}
    \caption{Inverse temperature dependence of the specific heat obtained from the RPPS approach (blue dashed-dotted line) and the RPMPS+T approach (orange dashed line) with 1024 samples in the 100-site Heisenberg chain. The upper panel represents the specific heat obtained from the fluctuation of the Hamiltonian [Eq.~\eqref{eq:C_fluctuation}] and the lower panel does that from the inverse temperature derivative of the thermal entropy [Eq.~\eqref{eq:C_entropy}]. Shaded regions indicate one-sigma uncertainty estimated from the jackknife analysis.\label{fig:specific_heat}}
\end{figure}

Figure~\ref{fig:specific_heat} represents the inverse temperature dependence of the specific heat obtained by the RPPS approach, the RPMPS+T approach, and the purification approach.
A standard way to compute the specific heat is to use the relation with the fluctuation of the total energy,
\begin{align}
    \label{eq:C_fluctuation}
    C_\mathrm{f} = k_\mathrm{B} \beta^2 (\braket{\hat{H}^2}_\beta - \braket{\hat{H}}^2_\beta).
\end{align}
This expression, however, is not useful in MPS simulations because the factor \(\beta^2\) enhances the truncation error as shown in the upper panel of Fig.~\ref{fig:specific_heat}~\footnote{We confirm that the increase of the specific heat at low temperatures is absent when the truncation error is reduced to \(10^{-10}\).}.
Instead, we use another expression of the specific heat
\begin{align}
    \label{eq:C_entropy}
    C_S = -\beta \frac{\partial S_\mathrm{th}}{\partial \beta}
\end{align}
which contains the factor \(\beta \) instead of \(\beta^2\).
The specific heat obtained from the inverse temperature derivative of the thermal entropy is presented in the lower panel of Fig.~\ref{fig:specific_heat}.
Here, the derivative is approximated by the second order central difference with finite difference \(\Delta \beta = 0.5J^{-1}_1\).
Likewise the partition function and thermal entropy, the values given by the RPMPS+T approach and the purification approach match and those from the RPPS approach deviates from them.
The deviation cannot be accounted for by the statistical error especially in \(5.0 \lesssim \beta J^{-1}_1\lesssim 8.0\) and around \(\beta J^{-1}_1 =20.0\).

\subsection{Frustrated zigzag chain}
In the previous subsection we analyzed the simple Heisenberg chain (\(J_2 = 0\)) in order to show that while the RPPS approach fails to reproduce the thermodynamic quantities because of sampling inefficiency, the RPMPS+T approach succeeds in reproducing them.
In this subsection, we examine the applicability of the RPMPS+T approach to frustrated systems, in which quantum Monte Carlo simulations suffer from negative-sign problems and applicable quasi-exact numerical approaches are severely limited.
Specifically, we choose the 100-site zigzag chain with \(J_2 = J_1\).
This system has a finite dimer gap when the magnetic field is absent~\cite{okunishi_magnetic_2003,okunishi_calculation_2008}.
In order to make simulations more challenging to emphasize the performances of the RPMPS+T approach, we simulate the canonical ensemble with the magnetization \(\sum_i \braket{\hat{S}^z_i}=10\), in which even the purification approach with a few thousands bond dimensions fails.

\begin{figure}
    \includegraphics[width=\linewidth]{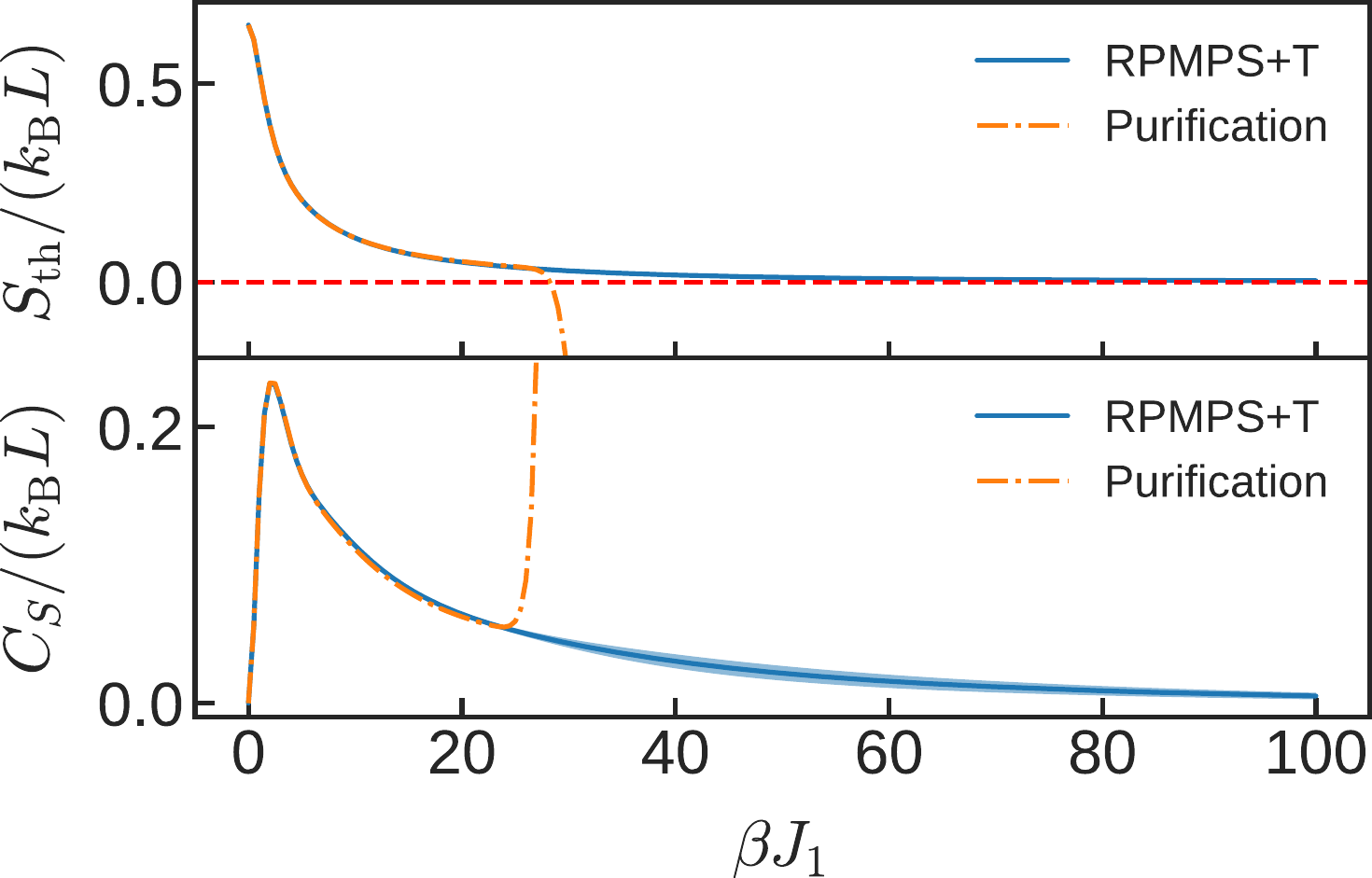}
    \caption{Inverse temperature dependence of the thermal entropy (the upper panel) and the specific heat of Eq.~\eqref{eq:C_entropy} (the lower panel) obtained by the RPMPS+T (blue solid line) approach and the purification approach (orange dashed-dotted line) in the 100-site zigzag chain. The number of samples used in the RPMPS+T approach is 512. A shaded region indicates the one-sigma uncertainty of the RPMPS+T results estimated from the jackknife analysis. The red dashed line denotes the physically possible lowest thermal entropy.\label{fig:SandC}}
\end{figure}

The upper panel of Fig.~\ref{fig:SandC} represents the thermal entropy of the 100-site zigzag chain obtained by the RPMPS+T approach and the purification approach.
From high to relatively low temperatures, namely, \(0 \lesssim \beta J_{1} \lesssim 20\), the thermal entropy given by the purification approach coincides with that given by the RPMPS+T approach.
However, when \(\beta \) increases in a low-temperature region \(\beta J_{1} \gtrsim 20\), the thermal entropy given by the purification approach rapidly decreases and eventually becomes negative.
Moreover, as shown in the lower panel of Fig.~\ref{fig:SandC}, the rapid growth of the specific heat obtained from the purification approach reflects the unphysical behavior of the thermal entropy.
These results manifest that the purification approach fails to describe the system even qualitatively.
On the contrary, the RPMPS+T approach gives the thermal entropy and specific heat which converge to zero at low temperatures as usual.
Therefore, the RPMPS+T approach can be applied to the spin 1/2 system in which even the purification approach fails.
The Trotter gate used in the RPMPS+T approach is the same one used in the performance tests of the Heisenberg chain, \((\Delta, n) = (5.0, 1)\). This success of the RPMPS+T approach in the zigzag chain also suggests that effective Trotter gates are reusable to some extent.

\begin{figure}
    \includegraphics[width=\linewidth]{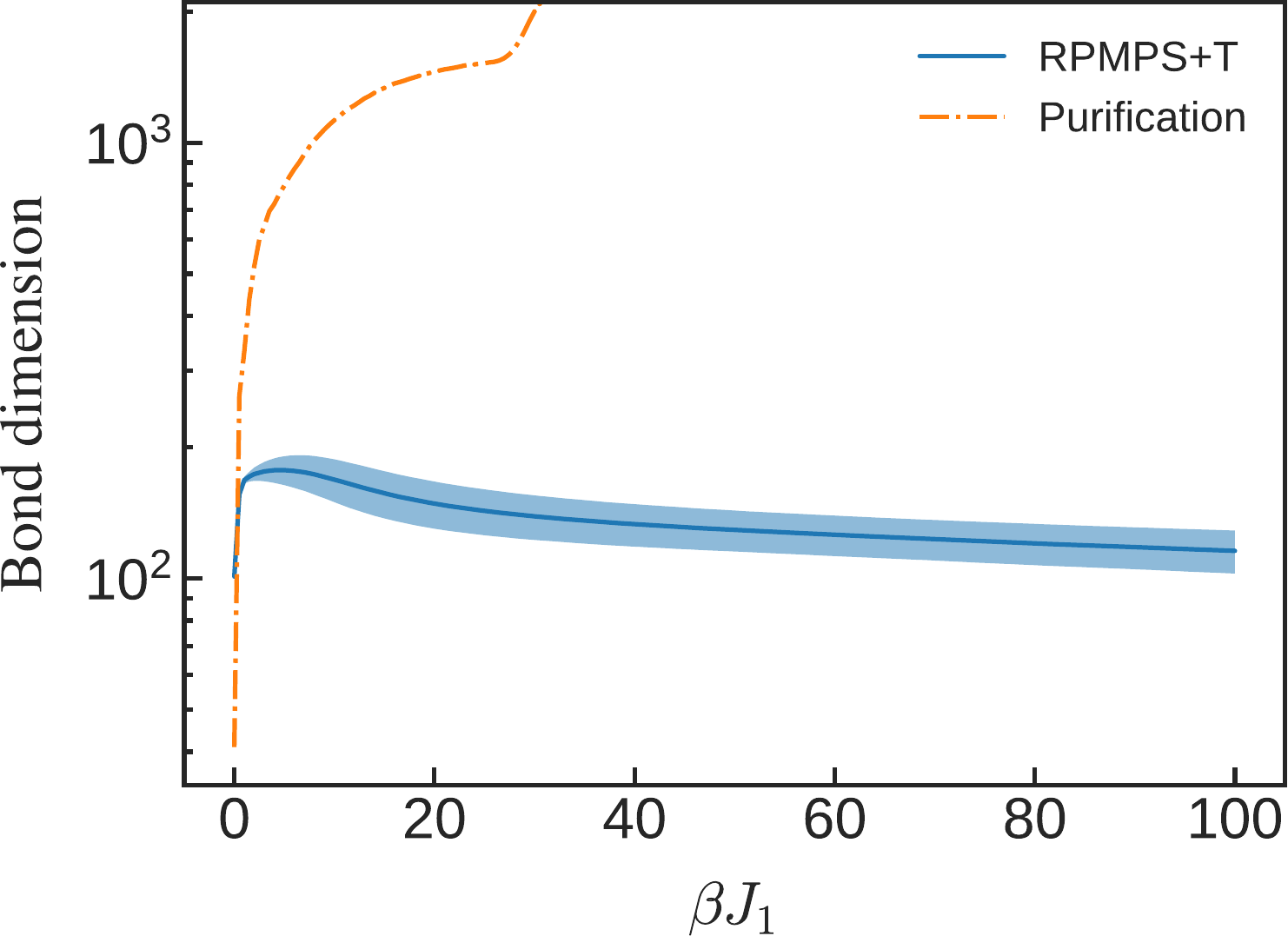}
    \caption{Inverse temperature dependence of the largest bond dimensions of MPS in the RPMPS+T approach (blue solid line) and the purification approach (orange dashed-dotted line) in the 100-site zigzag chain. The number of samples used in the RPMPS+T approach is 512. A shaded region indicates the standard deviation of the largest bond dimensions in the RPMPS+T approach.\label{fig:BondDim}}
\end{figure}

The failure of the purification approach stems from accumulated truncation errors during the imaginary-time evolution.
Nevertheless, reducing the truncation error for one operation from \(10^{-8}\) is not a practical solution because the largest bond dimension has already exceeded 1000 as shown in Fig.~\ref{fig:BondDim}, which represents the largest bond dimensions of MPS in the RPMPS+T approach and the purification approach.
It will require massive numerical cost to simulate this system properly with the purification approach.
On the other hand, in the RPMPS+T approach, the simulation is robust against the truncation  error and the bond dimensions of MPS in most of samples are less than 200.
Therefore, this approach has room to simulate the system more precisely, and to be applied to more complicated systems.
 
\section{Discussion and summaries\label{sec:summaries}}
\begin{figure}
    \includegraphics[width=\linewidth]{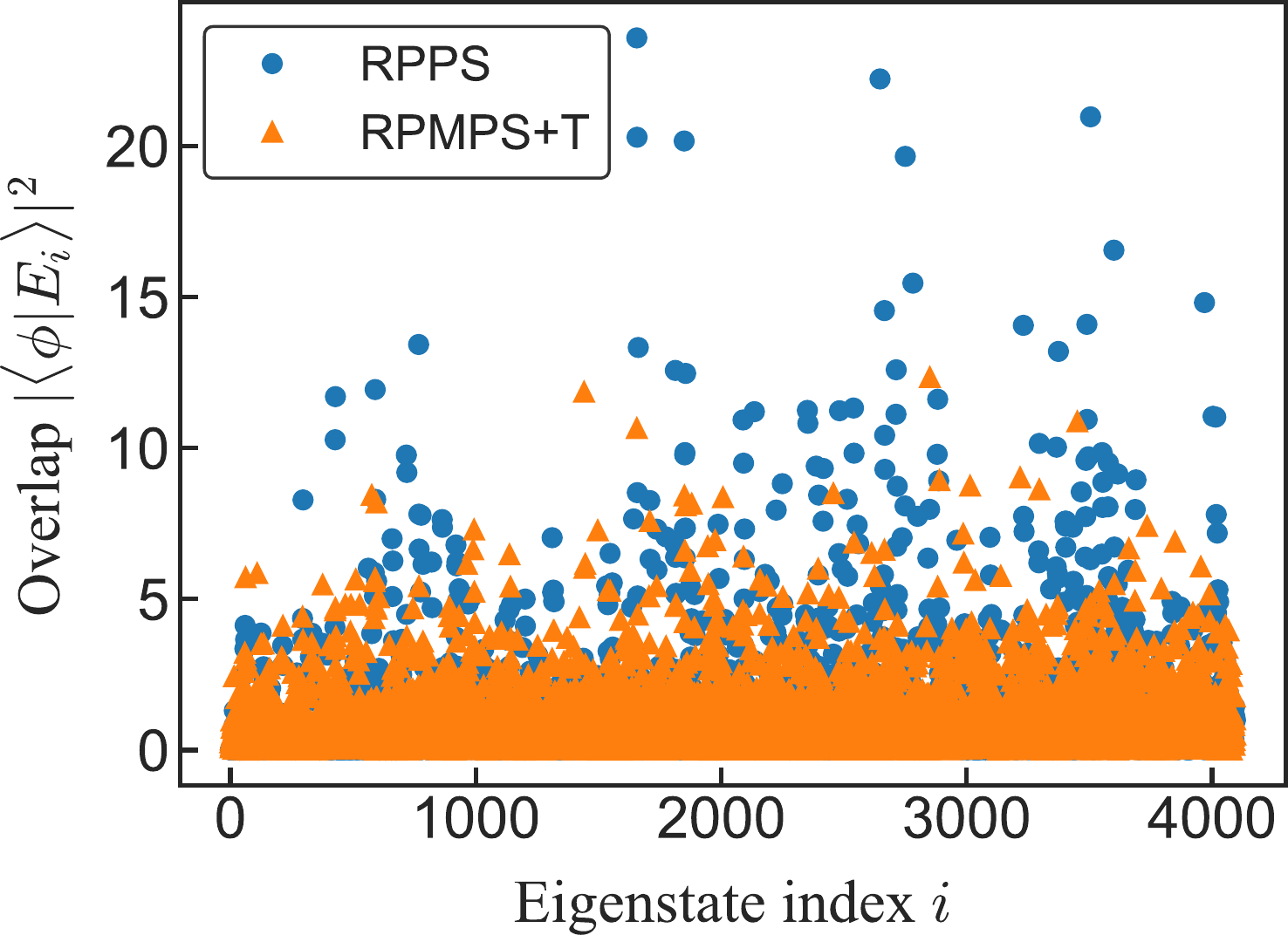}
    \caption{Single shot overlaps of eigenstates of the system Hamiltonian with initial states \(\ket{\phi}\) for the RPPS approach (blue circle) and the RPMPS+T approach (orange triangle) in the 12-site Heisenberg chain. We use the same RPPS for both approaches and \((\Delta, n) = (5.0, 1)\) for the parameters of the Trotter gate.\label{fig:dist}}
\end{figure}
Why do Trotter gates increase the sampling efficiency?
An answer to the question is that Trotter gates can make flatter the distribution of eigenstates of the system Hamiltonian in initial states.
If unnormalized initial states \(\ket{\psi_\mathrm{ini}}\) are superposition of all the equally weighted eigenstates \(\ket{E_i}\), namely \(\ket{\psi_\mathrm{ini}} = \sum_i\mathrm{e}^{i \theta_i}\ket{E_i}\), the quantity \(\braket{\psi_\mathrm{ini}|\mathrm{e}^{-\beta \hat{H}}|\psi_\mathrm{ini}}\) at each sample always gives the exact value of the partition function. In other words, the ideal sampling is achieved.
This indicates that the flatter the distribution of eigenstates is, the more efficient the sampling is.
In order to demonstrate that the Trotter gates make the distribution significantly flatter, we obtain all the eigenstates of the 12-site Heisenberg chain \(\ket{E_i}\) and calculate the overlap of the eigenstate with an initial state \(|\braket{E_i|p}|^2\) for the RPPS approach and \(|\braket{E_i|\hat{U}|p}|^2\) for the RPMPS+T approach with the same RPPS \(\ket{p}\).
For the parameters of the Trotter gates, we use \((\Delta, n) = (5.0, 1)\).
Figure~\ref{fig:dist} shows the single shot overlaps of the eigenstates with initial states \(\ket{\phi}\) for the RPPS and RPMPS+T approaches.
Compared to the distribution for the RPPS approach, that for the RPMPS+T approach is noticeably closer to the ideal flat distribution, \(|\braket{\phi|E_i}|^2 = 1\).

From the overlap of an eigenstate with an initial state, one can judge whether the number of samples is large enough to reproduce the statistics of the sampling.
Initial states \(\ket{\phi}\) should fulfill \(\overline{|\braket{\phi|E_i}|^2} = 1\) for all eigenstates \(\ket{E_i}\) since \(\overline{\braket{\phi|\mathrm{e}^{-\beta \hat{H}}|\phi}} = \sum_i \overline{|\braket{\phi|E_i}|^2} \mathrm{e}^{-\beta E_i}\) coincides with the partition function for any inverse temperature \(\beta \).
Here, \(E_i\) is the eigenenergy of an eigenstate \(\ket{E_i}\).
For large-scale systems, it is not feasible to obtain all of the eigenstates but one can access the ground state \(\ket{E_0}\) in (quasi-)1D systems by the DMRG algorithm.
Therefore, one can check the necessary condition \(\overline{|\braket{\phi|E_0}|^2} = 1\) even in large-scale systems.
With the same 1024 RPPS in the 100-site Heisenberg chain, \(\overline{|\braket{\phi|E_0}|^2}\) is \(1.04 \pm 0.18\) for the RPMPS+T approach and \((1.37 \pm 0.77) \times 10^{-2}\) for the RPPS approach.
From these values, one can confirm that the RPMPS+T approach with 1024 samples correctly reproduce the statistics and that the RPPS approach fails to reproduce it.
The deviations of thermodynamic quantities in the RPPS approach observed in Figs.~\ref{fig:ZandS} and~\ref{fig:specific_heat} are the results of this failure.
We also confirm that the less efficient RPMPS+T approach with the Trotter gates characterized by \((\Delta, n) = (10.0, 1)\) gives \(\overline{|\braket{\phi|E_0}|^2} = 2.81 \pm 1.83\) for the same 1024 RPPS.
The less efficient RPMPS+T approach fulfills the necessary condition within the statistical error, but the error is quite larger than that of the efficient one.
For finding good Trotter gates, the assessment based on the overlap with the ground state would be useful since this scheme does not require imaginary-time evolution for each sample.

The RPMPS+T approach has some advantages over existing approaches for simulating quantum many-body systems at finite temperatures via the sampling of pure states.
One of the most famous such approaches is the METTS algorithm~\cite{white_minimally_2009,stoudenmire_minimally_2010}.
The METTS algorithm has the importance sampling scheme implemented via the Markov chain.
An obvious advantage of the RPMPS+T approach is the absence of the importance sampling, which enables efficient parallel computations.
Another advantage is that one can compute the thermal averages of observables in a wide range of temperature from a set of samples while a set of samples in a single Markov chain allows for those at a single temperature.

Recently, \textcite{iwaki_thermal_2021} have introduced the TPQ MPS approach which does not possess the importance sampling scheme either.
They also use initially entangled states so that the resolution of the sampling efficiency in the TPQ MPS approach may be attributed to the use of entanglement as in the RPMPS+T approach.

A notable advantage is the compatibility with the Abelian symmetries.
The utilization of Abelian symmetries allows us to access the canonical ensemble.
It is crucial if one would like to simulate quasi-1D systems or dynamical behaviors.
The benefits of the Abelian symmetries become relevant in numerically challenging situations.
Besides, if the partition function of every symmetric sector is obtained, one can construct the grand canonical partition function with any magnetic field \(h\) at the inverse temperature \(\beta \) by
\begin{align}
    \Xi = \sum^{L/2}_{m=-L/2} \mathrm{e}^{\beta hm} Z_m(\beta).
\end{align}
Here, \(Z_m(\beta)\) is the partition function of the canonical ensemble without magnetic field in the symmetric sector corresponding to magnetization \(m\).

In summary, we developed a numerical approach using matrix product states to simulate quantum many-body systems at finite temperatures.
We improved the random phase product state (RPPS) approach by starting from an entangled state generated by the operation of the Trotter gates.
We defined indicators to quantify the sampling efficiency and assessed the efficiency of the RPPS approach and the improved approach in the Heisenberg chain.
We confirmed that the efficiency of the improved approach is higher than the RPPS approach by an order of magnitude.
Besides, we showed that the improved approach reproduces thermodynamic quantities consistent with those given by the purification approach while the RPPS gives unphysical negative thermal entropy.
We also demonstrated that the improved approach can simulate the thermodynamics of a frustrated zigzag chain in which even the purification approach fails.

Our improved approach serves as a valid option for investigating thermodynamics of quantum many-body systems in one dimension.
Moreover, it will be meaningful to further extend the improved approach to nonequilibrium dynamics.
Such an extension will be of immediate interest in the context of ultracold gases in optical lattices, in which much attention has been paid to nonequilibrium dynamics of one-dimensional quantum many-body systems at finite temperatures~\cite{greif_short-range_2013,tanzi_velocity-dependent_2016,yang_cooling_2020,taie_observation_2020}.
 
\begin{acknowledgments}
We thank A.~Iwaki and C.~Hotta for fruitful discussions.
The MPS calculations in this work are performed with ITensor library~\cite{fishman_itensor_2020}.
The reference C++ codes of the RPMPS+T approach are available at the github repository \url{https://github.com/ShimpeiGoto/RPMPS-T}.
This work was financially supported by KAKENHI from Japan Society for Promotion of Science: Grant No.\ 18K03492, No.\ 18H05228, and No.\ 20K14377, by CREST, JST No.\ JPMJCR1673, and by MEXT Q-LEAP Grant No.\ JPMXS0118069021.
\end{acknowledgments}


\begin{thebibliography}{43}%
\makeatletter
\providecommand \@ifxundefined [1]{%
 \@ifx{#1\undefined}
}%
\providecommand \@ifnum [1]{%
 \ifnum #1\expandafter \@firstoftwo
 \else \expandafter \@secondoftwo
 \fi
}%
\providecommand \@ifx [1]{%
 \ifx #1\expandafter \@firstoftwo
 \else \expandafter \@secondoftwo
 \fi
}%
\providecommand \natexlab [1]{#1}%
\providecommand \enquote  [1]{``#1''}%
\providecommand \bibnamefont  [1]{#1}%
\providecommand \bibfnamefont [1]{#1}%
\providecommand \citenamefont [1]{#1}%
\providecommand \href@noop [0]{\@secondoftwo}%
\providecommand \href [0]{\begingroup \@sanitize@url \@href}%
\providecommand \@href[1]{\@@startlink{#1}\@@href}%
\providecommand \@@href[1]{\endgroup#1\@@endlink}%
\providecommand \@sanitize@url [0]{\catcode `\\12\catcode `\$12\catcode
  `\&12\catcode `\#12\catcode `\^12\catcode `\_12\catcode `\%12\relax}%
\providecommand \@@startlink[1]{}%
\providecommand \@@endlink[0]{}%
\providecommand \url  [0]{\begingroup\@sanitize@url \@url }%
\providecommand \@url [1]{\endgroup\@href {#1}{\urlprefix }}%
\providecommand \urlprefix  [0]{URL }%
\providecommand \Eprint [0]{\href }%
\providecommand \doibase [0]{https://doi.org/}%
\providecommand \selectlanguage [0]{\@gobble}%
\providecommand \bibinfo  [0]{\@secondoftwo}%
\providecommand \bibfield  [0]{\@secondoftwo}%
\providecommand \translation [1]{[#1]}%
\providecommand \BibitemOpen [0]{}%
\providecommand \bibitemStop [0]{}%
\providecommand \bibitemNoStop [0]{.\EOS\space}%
\providecommand \EOS [0]{\spacefactor3000\relax}%
\providecommand \BibitemShut  [1]{\csname bibitem#1\endcsname}%
\let\auto@bib@innerbib\@empty
%</preamble>
\bibitem [{\citenamefont
  {Schollw{\"o}ck}(2011)}]{schollwock_density-matrix_2011}%
  \BibitemOpen
  \bibfield  {author} {\bibinfo {author} {\bibfnamefont {U.}~\bibnamefont
  {Schollw{\"o}ck}},\ }\href {https://doi.org/10.1016/j.aop.2010.09.012}
  {\bibfield  {journal} {\bibinfo  {journal} {Annals of Physics}\ }\textbf
  {\bibinfo {volume} {326}},\ \bibinfo {pages} {96} (\bibinfo {year}
  {2011})}\BibitemShut {NoStop}%
\bibitem [{\citenamefont {Hastings}(2007{\natexlab{a}})}]{hastings_area_2007}%
  \BibitemOpen
  \bibfield  {author} {\bibinfo {author} {\bibfnamefont {M.~B.}\ \bibnamefont
  {Hastings}},\ }\href {https://doi.org/10.1088/1742-5468/2007/08/P08024}
  {\bibfield  {journal} {\bibinfo  {journal} {J. Stat. Mech.}\ }\textbf
  {\bibinfo {volume} {2007}},\ \bibinfo {pages} {P08024} (\bibinfo {year}
  {2007}{\natexlab{a}})}\BibitemShut {NoStop}%
\bibitem [{\citenamefont
  {Hastings}(2007{\natexlab{b}})}]{hastings_entropy_2007}%
  \BibitemOpen
  \bibfield  {author} {\bibinfo {author} {\bibfnamefont {M.~B.}\ \bibnamefont
  {Hastings}},\ }\href {https://doi.org/10.1103/PhysRevB.76.035114} {\bibfield
  {journal} {\bibinfo  {journal} {Phys. Rev. B}\ }\textbf {\bibinfo {volume}
  {76}},\ \bibinfo {pages} {035114} (\bibinfo {year}
  {2007}{\natexlab{b}})}\BibitemShut {NoStop}%
\bibitem [{\citenamefont {Eisert}\ \emph {et~al.}(2010)\citenamefont {Eisert},
  \citenamefont {Cramer},\ and\ \citenamefont
  {Plenio}}]{eisert_textitcolloquium_2010}%
  \BibitemOpen
  \bibfield  {author} {\bibinfo {author} {\bibfnamefont {J.}~\bibnamefont
  {Eisert}}, \bibinfo {author} {\bibfnamefont {M.}~\bibnamefont {Cramer}},\
  and\ \bibinfo {author} {\bibfnamefont {M.~B.}\ \bibnamefont {Plenio}},\
  }\href {https://doi.org/10.1103/RevModPhys.82.277} {\bibfield  {journal}
  {\bibinfo  {journal} {Rev. Mod. Phys.}\ }\textbf {\bibinfo {volume} {82}},\
  \bibinfo {pages} {277} (\bibinfo {year} {2010})}\BibitemShut {NoStop}%
\bibitem [{\citenamefont {White}(1992)}]{white_density_1992}%
  \BibitemOpen
  \bibfield  {author} {\bibinfo {author} {\bibfnamefont {S.~R.}\ \bibnamefont
  {White}},\ }\href {https://doi.org/10.1103/PhysRevLett.69.2863} {\bibfield
  {journal} {\bibinfo  {journal} {Phys. Rev. Lett.}\ }\textbf {\bibinfo
  {volume} {69}},\ \bibinfo {pages} {2863} (\bibinfo {year}
  {1992})}\BibitemShut {NoStop}%
\bibitem [{\citenamefont {White}(1993)}]{white_density-matrix_1993}%
  \BibitemOpen
  \bibfield  {author} {\bibinfo {author} {\bibfnamefont {S.~R.}\ \bibnamefont
  {White}},\ }\href {https://doi.org/10.1103/PhysRevB.48.10345} {\bibfield
  {journal} {\bibinfo  {journal} {Phys. Rev. B}\ }\textbf {\bibinfo {volume}
  {48}},\ \bibinfo {pages} {10345} (\bibinfo {year} {1993})}\BibitemShut
  {NoStop}%
\bibitem [{\citenamefont {Khemani}\ \emph {et~al.}(2016)\citenamefont
  {Khemani}, \citenamefont {Pollmann},\ and\ \citenamefont
  {Sondhi}}]{khemani_obtaining_2016}%
  \BibitemOpen
  \bibfield  {author} {\bibinfo {author} {\bibfnamefont {V.}~\bibnamefont
  {Khemani}}, \bibinfo {author} {\bibfnamefont {F.}~\bibnamefont {Pollmann}},\
  and\ \bibinfo {author} {\bibfnamefont {S.~L.}\ \bibnamefont {Sondhi}},\
  }\href {https://doi.org/10.1103/PhysRevLett.116.247204} {\bibfield  {journal}
  {\bibinfo  {journal} {Phys. Rev. Lett.}\ }\textbf {\bibinfo {volume} {116}},\
  \bibinfo {pages} {247204} (\bibinfo {year} {2016})}\BibitemShut {NoStop}%
\bibitem [{\citenamefont {Vidal}(2003)}]{vidal_efficient_2003}%
  \BibitemOpen
  \bibfield  {author} {\bibinfo {author} {\bibfnamefont {G.}~\bibnamefont
  {Vidal}},\ }\href {https://doi.org/10.1103/PhysRevLett.91.147902} {\bibfield
  {journal} {\bibinfo  {journal} {Phys. Rev. Lett.}\ }\textbf {\bibinfo
  {volume} {91}},\ \bibinfo {pages} {147902} (\bibinfo {year}
  {2003})}\BibitemShut {NoStop}%
\bibitem [{\citenamefont {Vidal}(2004)}]{vidal_efficient_2004}%
  \BibitemOpen
  \bibfield  {author} {\bibinfo {author} {\bibfnamefont {G.}~\bibnamefont
  {Vidal}},\ }\href {https://doi.org/10.1103/PhysRevLett.93.040502} {\bibfield
  {journal} {\bibinfo  {journal} {Phys. Rev. Lett.}\ }\textbf {\bibinfo
  {volume} {93}},\ \bibinfo {pages} {040502} (\bibinfo {year}
  {2004})}\BibitemShut {NoStop}%
\bibitem [{\citenamefont {Daley}\ \emph {et~al.}(2004)\citenamefont {Daley},
  \citenamefont {Kollath}, \citenamefont {Schollw{\"o}ck},\ and\ \citenamefont
  {Vidal}}]{daley_time-dependent_2004}%
  \BibitemOpen
  \bibfield  {author} {\bibinfo {author} {\bibfnamefont {A.~J.}\ \bibnamefont
  {Daley}}, \bibinfo {author} {\bibfnamefont {C.}~\bibnamefont {Kollath}},
  \bibinfo {author} {\bibfnamefont {U.}~\bibnamefont {Schollw{\"o}ck}},\ and\
  \bibinfo {author} {\bibfnamefont {G.}~\bibnamefont {Vidal}},\ }\href
  {https://doi.org/10.1088/1742-5468/2004/04/P04005} {\bibfield  {journal}
  {\bibinfo  {journal} {J. Stat. Mech.}\ }\textbf {\bibinfo {volume} {2004}},\
  \bibinfo {pages} {P04005} (\bibinfo {year} {2004})}\BibitemShut {NoStop}%
\bibitem [{\citenamefont {White}\ and\ \citenamefont
  {Feiguin}(2004)}]{white_real-time_2004}%
  \BibitemOpen
  \bibfield  {author} {\bibinfo {author} {\bibfnamefont {S.~R.}\ \bibnamefont
  {White}}\ and\ \bibinfo {author} {\bibfnamefont {A.~E.}\ \bibnamefont
  {Feiguin}},\ }\href {https://doi.org/10.1103/PhysRevLett.93.076401}
  {\bibfield  {journal} {\bibinfo  {journal} {Phys. Rev. Lett.}\ }\textbf
  {\bibinfo {volume} {93}},\ \bibinfo {pages} {076401} (\bibinfo {year}
  {2004})}\BibitemShut {NoStop}%
\bibitem [{\citenamefont {Zwolak}\ and\ \citenamefont
  {Vidal}(2004)}]{zwolak_mixed-state_2004}%
  \BibitemOpen
  \bibfield  {author} {\bibinfo {author} {\bibfnamefont {M.}~\bibnamefont
  {Zwolak}}\ and\ \bibinfo {author} {\bibfnamefont {G.}~\bibnamefont {Vidal}},\
  }\href {https://doi.org/10.1103/PhysRevLett.93.207205} {\bibfield  {journal}
  {\bibinfo  {journal} {Phys. Rev. Lett.}\ }\textbf {\bibinfo {volume} {93}},\
  \bibinfo {pages} {207205} (\bibinfo {year} {2004})}\BibitemShut {NoStop}%
\bibitem [{\citenamefont {Verstraete}\ \emph {et~al.}(2004)\citenamefont
  {Verstraete}, \citenamefont {{Garc{\'i}a-Ripoll}},\ and\ \citenamefont
  {Cirac}}]{verstraete_matrix_2004}%
  \BibitemOpen
  \bibfield  {author} {\bibinfo {author} {\bibfnamefont {F.}~\bibnamefont
  {Verstraete}}, \bibinfo {author} {\bibfnamefont {J.~J.}\ \bibnamefont
  {{Garc{\'i}a-Ripoll}}},\ and\ \bibinfo {author} {\bibfnamefont {J.~I.}\
  \bibnamefont {Cirac}},\ }\href
  {https://doi.org/10.1103/PhysRevLett.93.207204} {\bibfield  {journal}
  {\bibinfo  {journal} {Phys. Rev. Lett.}\ }\textbf {\bibinfo {volume} {93}},\
  \bibinfo {pages} {207204} (\bibinfo {year} {2004})}\BibitemShut {NoStop}%
\bibitem [{\citenamefont {Feiguin}\ and\ \citenamefont
  {White}(2005)}]{feiguin_finite-temperature_2005}%
  \BibitemOpen
  \bibfield  {author} {\bibinfo {author} {\bibfnamefont {A.~E.}\ \bibnamefont
  {Feiguin}}\ and\ \bibinfo {author} {\bibfnamefont {S.~R.}\ \bibnamefont
  {White}},\ }\href {https://doi.org/10.1103/PhysRevB.72.220401} {\bibfield
  {journal} {\bibinfo  {journal} {Phys. Rev. B}\ }\textbf {\bibinfo {volume}
  {72}},\ \bibinfo {pages} {220401} (\bibinfo {year} {2005})}\BibitemShut
  {NoStop}%
\bibitem [{\citenamefont {White}(2009)}]{white_minimally_2009}%
  \BibitemOpen
  \bibfield  {author} {\bibinfo {author} {\bibfnamefont {S.~R.}\ \bibnamefont
  {White}},\ }\href {https://doi.org/10.1103/PhysRevLett.102.190601} {\bibfield
   {journal} {\bibinfo  {journal} {Phys. Rev. Lett.}\ }\textbf {\bibinfo
  {volume} {102}},\ \bibinfo {pages} {190601} (\bibinfo {year}
  {2009})}\BibitemShut {NoStop}%
\bibitem [{\citenamefont {Stoudenmire}\ and\ \citenamefont
  {White}(2010)}]{stoudenmire_minimally_2010}%
  \BibitemOpen
  \bibfield  {author} {\bibinfo {author} {\bibfnamefont {E.~M.}\ \bibnamefont
  {Stoudenmire}}\ and\ \bibinfo {author} {\bibfnamefont {S.~R.}\ \bibnamefont
  {White}},\ }\href {https://doi.org/10.1088/1367-2630/12/5/055026} {\bibfield
  {journal} {\bibinfo  {journal} {New J. Phys.}\ }\textbf {\bibinfo {volume}
  {12}},\ \bibinfo {pages} {055026} (\bibinfo {year} {2010})}\BibitemShut
  {NoStop}%
\bibitem [{\citenamefont {{\L}{\k{a}}cki}\ \emph {et~al.}(2015)\citenamefont
  {{\L}{\k{a}}cki}, \citenamefont {Delande},\ and\ \citenamefont
  {Zakrzewski}}]{lacki_dynamics_2015}%
  \BibitemOpen
  \bibfield  {author} {\bibinfo {author} {\bibfnamefont {M.}~\bibnamefont
  {{\L}{\k{a}}cki}}, \bibinfo {author} {\bibfnamefont {D.}~\bibnamefont
  {Delande}},\ and\ \bibinfo {author} {\bibfnamefont {J.}~\bibnamefont
  {Zakrzewski}},\ }\href {https://doi.org/10.1103/PhysRevB.91.134304}
  {\bibfield  {journal} {\bibinfo  {journal} {Phys. Rev. B}\ }\textbf {\bibinfo
  {volume} {91}},\ \bibinfo {pages} {134304} (\bibinfo {year}
  {2015})}\BibitemShut {NoStop}%
\bibitem [{\citenamefont {Binder}\ and\ \citenamefont
  {Barthel}(2017)}]{binder_symmetric_2017}%
  \BibitemOpen
  \bibfield  {author} {\bibinfo {author} {\bibfnamefont {M.}~\bibnamefont
  {Binder}}\ and\ \bibinfo {author} {\bibfnamefont {T.}~\bibnamefont
  {Barthel}},\ }\href {https://doi.org/10.1103/PhysRevB.95.195148} {\bibfield
  {journal} {\bibinfo  {journal} {Phys. Rev. B}\ }\textbf {\bibinfo {volume}
  {95}},\ \bibinfo {pages} {195148} (\bibinfo {year} {2017})}\BibitemShut
  {NoStop}%
\bibitem [{\citenamefont {Chung}\ and\ \citenamefont
  {Schollw{\"o}ck}(2019)}]{chung_minimally_2019}%
  \BibitemOpen
  \bibfield  {author} {\bibinfo {author} {\bibfnamefont {C.-M.}\ \bibnamefont
  {Chung}}\ and\ \bibinfo {author} {\bibfnamefont {U.}~\bibnamefont
  {Schollw{\"o}ck}},\ }\href@noop {} {\bibfield  {journal} {\bibinfo  {journal}
  {ArXiv191003329 Cond-Mat}\ } (\bibinfo {year} {2019})},\ \Eprint
  {https://arxiv.org/abs/1910.03329} {arXiv:1910.03329 [cond-mat]} \BibitemShut
  {NoStop}%
\bibitem [{\citenamefont {Chen}\ and\ \citenamefont
  {Stoudenmire}(2020)}]{chen_hybrid_2020}%
  \BibitemOpen
  \bibfield  {author} {\bibinfo {author} {\bibfnamefont {J.}~\bibnamefont
  {Chen}}\ and\ \bibinfo {author} {\bibfnamefont {E.~M.}\ \bibnamefont
  {Stoudenmire}},\ }\href {https://doi.org/10.1103/PhysRevB.101.195119}
  {\bibfield  {journal} {\bibinfo  {journal} {Phys. Rev. B}\ }\textbf {\bibinfo
  {volume} {101}},\ \bibinfo {pages} {195119} (\bibinfo {year}
  {2020})}\BibitemShut {NoStop}%
\bibitem [{\citenamefont {Goto}\ and\ \citenamefont
  {Danshita}(2019)}]{goto_quasiexact_2019}%
  \BibitemOpen
  \bibfield  {author} {\bibinfo {author} {\bibfnamefont {S.}~\bibnamefont
  {Goto}}\ and\ \bibinfo {author} {\bibfnamefont {I.}~\bibnamefont
  {Danshita}},\ }\href {https://doi.org/10.1103/PhysRevLett.123.143002}
  {\bibfield  {journal} {\bibinfo  {journal} {Phys. Rev. Lett.}\ }\textbf
  {\bibinfo {volume} {123}},\ \bibinfo {pages} {143002} (\bibinfo {year}
  {2019})}\BibitemShut {NoStop}%
\bibitem [{\citenamefont {Goto}\ and\ \citenamefont
  {Danshita}(2020)}]{goto_minimally_2020}%
  \BibitemOpen
  \bibfield  {author} {\bibinfo {author} {\bibfnamefont {S.}~\bibnamefont
  {Goto}}\ and\ \bibinfo {author} {\bibfnamefont {I.}~\bibnamefont
  {Danshita}},\ }\href {https://doi.org/10.1103/PhysRevResearch.2.043236}
  {\bibfield  {journal} {\bibinfo  {journal} {Phys. Rev. Research}\ }\textbf
  {\bibinfo {volume} {2}},\ \bibinfo {pages} {043236} (\bibinfo {year}
  {2020})}\BibitemShut {NoStop}%
\bibitem [{\citenamefont {Iwaki}\ \emph {et~al.}(2021)\citenamefont {Iwaki},
  \citenamefont {Shimizu},\ and\ \citenamefont {Hotta}}]{iwaki_thermal_2021}%
  \BibitemOpen
  \bibfield  {author} {\bibinfo {author} {\bibfnamefont {A.}~\bibnamefont
  {Iwaki}}, \bibinfo {author} {\bibfnamefont {A.}~\bibnamefont {Shimizu}},\
  and\ \bibinfo {author} {\bibfnamefont {C.}~\bibnamefont {Hotta}},\ }\href
  {https://doi.org/10.1103/PhysRevResearch.3.L022015} {\bibfield  {journal}
  {\bibinfo  {journal} {Phys. Rev. Research}\ }\textbf {\bibinfo {volume}
  {3}},\ \bibinfo {pages} {L022015} (\bibinfo {year} {2021})}\BibitemShut
  {NoStop}%
\bibitem [{\citenamefont {Iitaka}(2020)}]{iitaka_random_2020}%
  \BibitemOpen
  \bibfield  {author} {\bibinfo {author} {\bibfnamefont {T.}~\bibnamefont
  {Iitaka}},\ }\href@noop {} {\bibfield  {journal} {\bibinfo  {journal}
  {ArXiv200614459 Cond-Mat Physicsquant-Ph}\ } (\bibinfo {year} {2020})},\
  \Eprint {https://arxiv.org/abs/2006.14459} {arXiv:2006.14459 [cond-mat,
  physics:quant-ph]} \BibitemShut {NoStop}%
\bibitem [{\citenamefont {Imada}\ and\ \citenamefont
  {Takahashi}(1986)}]{imada_quantum_1986}%
  \BibitemOpen
  \bibfield  {author} {\bibinfo {author} {\bibfnamefont {M.}~\bibnamefont
  {Imada}}\ and\ \bibinfo {author} {\bibfnamefont {M.}~\bibnamefont
  {Takahashi}},\ }\href {https://doi.org/10.1143/JPSJ.55.3354} {\bibfield
  {journal} {\bibinfo  {journal} {J. Phys. Soc. Jpn.}\ }\textbf {\bibinfo
  {volume} {55}},\ \bibinfo {pages} {3354} (\bibinfo {year}
  {1986})}\BibitemShut {NoStop}%
\bibitem [{\citenamefont {Lloyd}(1988)}]{lloyd_pure_1988}%
  \BibitemOpen
  \bibfield  {author} {\bibinfo {author} {\bibfnamefont {S.}~\bibnamefont
  {Lloyd}},\ }\emph {\bibinfo {title} {Pure State Quantum Statistical Mechanics
  and Black Holes}},\ \href@noop {} {Ph.D. thesis},\ \bibinfo  {school}
  {Rockfeller University} (\bibinfo {year} {1988})\BibitemShut {NoStop}%
\bibitem [{\citenamefont {Jakli{\v c}}\ and\ \citenamefont {Prelov{\v
  s}ek}(1994)}]{jaklic_lanczos_1994}%
  \BibitemOpen
  \bibfield  {author} {\bibinfo {author} {\bibfnamefont {J.}~\bibnamefont
  {Jakli{\v c}}}\ and\ \bibinfo {author} {\bibfnamefont {P.}~\bibnamefont
  {Prelov{\v s}ek}},\ }\href {https://doi.org/10.1103/PhysRevB.49.5065}
  {\bibfield  {journal} {\bibinfo  {journal} {Phys. Rev. B}\ }\textbf {\bibinfo
  {volume} {49}},\ \bibinfo {pages} {5065} (\bibinfo {year}
  {1994})}\BibitemShut {NoStop}%
\bibitem [{\citenamefont {Hams}\ and\ \citenamefont
  {De~Raedt}(2000)}]{hams_fast_2000}%
  \BibitemOpen
  \bibfield  {author} {\bibinfo {author} {\bibfnamefont {A.}~\bibnamefont
  {Hams}}\ and\ \bibinfo {author} {\bibfnamefont {H.}~\bibnamefont
  {De~Raedt}},\ }\href {https://doi.org/10.1103/PhysRevE.62.4365} {\bibfield
  {journal} {\bibinfo  {journal} {Phys. Rev. E}\ }\textbf {\bibinfo {volume}
  {62}},\ \bibinfo {pages} {4365} (\bibinfo {year} {2000})}\BibitemShut
  {NoStop}%
\bibitem [{\citenamefont {Sugiura}\ and\ \citenamefont
  {Shimizu}(2012)}]{sugiura_thermal_2012}%
  \BibitemOpen
  \bibfield  {author} {\bibinfo {author} {\bibfnamefont {S.}~\bibnamefont
  {Sugiura}}\ and\ \bibinfo {author} {\bibfnamefont {A.}~\bibnamefont
  {Shimizu}},\ }\href {https://doi.org/10.1103/PhysRevLett.108.240401}
  {\bibfield  {journal} {\bibinfo  {journal} {Phys. Rev. Lett.}\ }\textbf
  {\bibinfo {volume} {108}},\ \bibinfo {pages} {240401} (\bibinfo {year}
  {2012})}\BibitemShut {NoStop}%
\bibitem [{\citenamefont {Sugiura}\ and\ \citenamefont
  {Shimizu}(2013)}]{sugiura_canonical_2013}%
  \BibitemOpen
  \bibfield  {author} {\bibinfo {author} {\bibfnamefont {S.}~\bibnamefont
  {Sugiura}}\ and\ \bibinfo {author} {\bibfnamefont {A.}~\bibnamefont
  {Shimizu}},\ }\href {https://doi.org/10.1103/PhysRevLett.111.010401}
  {\bibfield  {journal} {\bibinfo  {journal} {Phys. Rev. Lett.}\ }\textbf
  {\bibinfo {volume} {111}},\ \bibinfo {pages} {010401} (\bibinfo {year}
  {2013})}\BibitemShut {NoStop}%
\bibitem [{\citenamefont {Nakagawa}\ \emph {et~al.}(2018)\citenamefont
  {Nakagawa}, \citenamefont {Watanabe}, \citenamefont {Fujita},\ and\
  \citenamefont {Sugiura}}]{nakagawa_universality_2018}%
  \BibitemOpen
  \bibfield  {author} {\bibinfo {author} {\bibfnamefont {Y.~O.}\ \bibnamefont
  {Nakagawa}}, \bibinfo {author} {\bibfnamefont {M.}~\bibnamefont {Watanabe}},
  \bibinfo {author} {\bibfnamefont {H.}~\bibnamefont {Fujita}},\ and\ \bibinfo
  {author} {\bibfnamefont {S.}~\bibnamefont {Sugiura}},\ }\href
  {https://doi.org/10.1038/s41467-018-03883-9} {\bibfield  {journal} {\bibinfo
  {journal} {Nat Commun}\ }\textbf {\bibinfo {volume} {9}},\ \bibinfo {pages}
  {1} (\bibinfo {year} {2018})}\BibitemShut {NoStop}%
\bibitem [{\citenamefont {Okunishi}\ and\ \citenamefont
  {Tonegawa}(2003)}]{okunishi_magnetic_2003}%
  \BibitemOpen
  \bibfield  {author} {\bibinfo {author} {\bibfnamefont {K.}~\bibnamefont
  {Okunishi}}\ and\ \bibinfo {author} {\bibfnamefont {T.}~\bibnamefont
  {Tonegawa}},\ }\href {https://doi.org/10.1143/JPSJ.72.479} {\bibfield
  {journal} {\bibinfo  {journal} {J. Phys. Soc. Jpn.}\ }\textbf {\bibinfo
  {volume} {72}},\ \bibinfo {pages} {479} (\bibinfo {year} {2003})}\BibitemShut
  {NoStop}%
\bibitem [{\citenamefont {Okunishi}(2008)}]{okunishi_calculation_2008}%
  \BibitemOpen
  \bibfield  {author} {\bibinfo {author} {\bibfnamefont {K.}~\bibnamefont
  {Okunishi}},\ }\href {https://doi.org/10.1143/JPSJ.77.114004} {\bibfield
  {journal} {\bibinfo  {journal} {J. Phys. Soc. Jpn.}\ }\textbf {\bibinfo
  {volume} {77}},\ \bibinfo {pages} {114004} (\bibinfo {year}
  {2008})}\BibitemShut {NoStop}%
\bibitem [{\citenamefont {Hotta}\ and\ \citenamefont
  {Shibata}(2012)}]{hotta_grand_2012}%
  \BibitemOpen
  \bibfield  {author} {\bibinfo {author} {\bibfnamefont {C.}~\bibnamefont
  {Hotta}}\ and\ \bibinfo {author} {\bibfnamefont {N.}~\bibnamefont
  {Shibata}},\ }\href {https://doi.org/10.1103/PhysRevB.86.041108} {\bibfield
  {journal} {\bibinfo  {journal} {Phys. Rev. B}\ }\textbf {\bibinfo {volume}
  {86}},\ \bibinfo {pages} {041108} (\bibinfo {year} {2012})}\BibitemShut
  {NoStop}%
\bibitem [{\citenamefont {Iitaka}\ and\ \citenamefont
  {Ebisuzaki}(2004)}]{iitaka_random_2004}%
  \BibitemOpen
  \bibfield  {author} {\bibinfo {author} {\bibfnamefont {T.}~\bibnamefont
  {Iitaka}}\ and\ \bibinfo {author} {\bibfnamefont {T.}~\bibnamefont
  {Ebisuzaki}},\ }\href {https://doi.org/10.1103/PhysRevE.69.057701} {\bibfield
   {journal} {\bibinfo  {journal} {Phys. Rev. E}\ }\textbf {\bibinfo {volume}
  {69}},\ \bibinfo {pages} {057701} (\bibinfo {year} {2004})}\BibitemShut
  {NoStop}%
\bibitem [{\citenamefont {Barthel}(2016)}]{barthel_matrix_2016}%
  \BibitemOpen
  \bibfield  {author} {\bibinfo {author} {\bibfnamefont {T.}~\bibnamefont
  {Barthel}},\ }\href {https://doi.org/10.1103/PhysRevB.94.115157} {\bibfield
  {journal} {\bibinfo  {journal} {Phys. Rev. B}\ }\textbf {\bibinfo {volume}
  {94}},\ \bibinfo {pages} {115157} (\bibinfo {year} {2016})}\BibitemShut
  {NoStop}%
\bibitem [{\citenamefont {Li}\ \emph {et~al.}(2021)\citenamefont {Li},
  \citenamefont {Han}, \citenamefont {Sun}, \citenamefont {Shang},\ and\
  \citenamefont {Zhu}}]{li_verification_2021}%
  \BibitemOpen
  \bibfield  {author} {\bibinfo {author} {\bibfnamefont {Z.}~\bibnamefont
  {Li}}, \bibinfo {author} {\bibfnamefont {Y.-G.}\ \bibnamefont {Han}},
  \bibinfo {author} {\bibfnamefont {H.-F.}\ \bibnamefont {Sun}}, \bibinfo
  {author} {\bibfnamefont {J.}~\bibnamefont {Shang}},\ and\ \bibinfo {author}
  {\bibfnamefont {H.}~\bibnamefont {Zhu}},\ }\href
  {https://doi.org/10.1103/PhysRevA.103.022601} {\bibfield  {journal} {\bibinfo
   {journal} {Phys. Rev. A}\ }\textbf {\bibinfo {volume} {103}},\ \bibinfo
  {pages} {022601} (\bibinfo {year} {2021})}\BibitemShut {NoStop}%
\bibitem [{Note1()}]{Note1}%
  \BibitemOpen
  \bibinfo {note} {We confirm that the increase of the specific heat at low
  temperatures is absent when the truncation error is reduced to
  \(10^{-10}\).}\BibitemShut {Stop}%
\bibitem [{\citenamefont {Greif}\ \emph {et~al.}(2013)\citenamefont {Greif},
  \citenamefont {Uehlinger}, \citenamefont {Jotzu}, \citenamefont {Tarruell},\
  and\ \citenamefont {Esslinger}}]{greif_short-range_2013}%
  \BibitemOpen
  \bibfield  {author} {\bibinfo {author} {\bibfnamefont {D.}~\bibnamefont
  {Greif}}, \bibinfo {author} {\bibfnamefont {T.}~\bibnamefont {Uehlinger}},
  \bibinfo {author} {\bibfnamefont {G.}~\bibnamefont {Jotzu}}, \bibinfo
  {author} {\bibfnamefont {L.}~\bibnamefont {Tarruell}},\ and\ \bibinfo
  {author} {\bibfnamefont {T.}~\bibnamefont {Esslinger}},\ }\href
  {https://doi.org/10.1126/science.1236362} {\bibfield  {journal} {\bibinfo
  {journal} {Science}\ }\textbf {\bibinfo {volume} {340}},\ \bibinfo {pages}
  {1307} (\bibinfo {year} {2013})}\BibitemShut {NoStop}%
\bibitem [{\citenamefont {Tanzi}\ \emph {et~al.}(2016)\citenamefont {Tanzi},
  \citenamefont {Scaffidi~Abbate}, \citenamefont {Cataldini}, \citenamefont
  {Gori}, \citenamefont {Lucioni}, \citenamefont {Inguscio}, \citenamefont
  {Modugno},\ and\ \citenamefont {D'Errico}}]{tanzi_velocity-dependent_2016}%
  \BibitemOpen
  \bibfield  {author} {\bibinfo {author} {\bibfnamefont {L.}~\bibnamefont
  {Tanzi}}, \bibinfo {author} {\bibfnamefont {S.}~\bibnamefont
  {Scaffidi~Abbate}}, \bibinfo {author} {\bibfnamefont {F.}~\bibnamefont
  {Cataldini}}, \bibinfo {author} {\bibfnamefont {L.}~\bibnamefont {Gori}},
  \bibinfo {author} {\bibfnamefont {E.}~\bibnamefont {Lucioni}}, \bibinfo
  {author} {\bibfnamefont {M.}~\bibnamefont {Inguscio}}, \bibinfo {author}
  {\bibfnamefont {G.}~\bibnamefont {Modugno}},\ and\ \bibinfo {author}
  {\bibfnamefont {C.}~\bibnamefont {D'Errico}},\ }\href
  {https://doi.org/10.1038/srep25965} {\bibfield  {journal} {\bibinfo
  {journal} {Sci. Rep.}\ }\textbf {\bibinfo {volume} {6}},\ \bibinfo {pages}
  {25965} (\bibinfo {year} {2016})}\BibitemShut {NoStop}%
\bibitem [{\citenamefont {Yang}\ \emph {et~al.}(2020)\citenamefont {Yang},
  \citenamefont {Sun}, \citenamefont {Huang}, \citenamefont {Wang},
  \citenamefont {Deng}, \citenamefont {Dai}, \citenamefont {Yuan},\ and\
  \citenamefont {Pan}}]{yang_cooling_2020}%
  \BibitemOpen
  \bibfield  {author} {\bibinfo {author} {\bibfnamefont {B.}~\bibnamefont
  {Yang}}, \bibinfo {author} {\bibfnamefont {H.}~\bibnamefont {Sun}}, \bibinfo
  {author} {\bibfnamefont {C.-J.}\ \bibnamefont {Huang}}, \bibinfo {author}
  {\bibfnamefont {H.-Y.}\ \bibnamefont {Wang}}, \bibinfo {author}
  {\bibfnamefont {Y.}~\bibnamefont {Deng}}, \bibinfo {author} {\bibfnamefont
  {H.-N.}\ \bibnamefont {Dai}}, \bibinfo {author} {\bibfnamefont {Z.-S.}\
  \bibnamefont {Yuan}},\ and\ \bibinfo {author} {\bibfnamefont {J.-W.}\
  \bibnamefont {Pan}},\ }\href {https://doi.org/10.1126/science.aaz6801}
  {\bibfield  {journal} {\bibinfo  {journal} {Science}\ }\textbf {\bibinfo
  {volume} {369}},\ \bibinfo {pages} {550} (\bibinfo {year}
  {2020})}\BibitemShut {NoStop}%
\bibitem [{\citenamefont {Taie}\ \emph {et~al.}(2020)\citenamefont {Taie},
  \citenamefont {{Ibarra-Garc{\'i}a-Padilla}}, \citenamefont {Nishizawa},
  \citenamefont {Takasu}, \citenamefont {Kuno}, \citenamefont {Wei},
  \citenamefont {Scalettar}, \citenamefont {Hazzard},\ and\ \citenamefont
  {Takahashi}}]{taie_observation_2020}%
  \BibitemOpen
  \bibfield  {author} {\bibinfo {author} {\bibfnamefont {S.}~\bibnamefont
  {Taie}}, \bibinfo {author} {\bibfnamefont {E.}~\bibnamefont
  {{Ibarra-Garc{\'i}a-Padilla}}}, \bibinfo {author} {\bibfnamefont
  {N.}~\bibnamefont {Nishizawa}}, \bibinfo {author} {\bibfnamefont
  {Y.}~\bibnamefont {Takasu}}, \bibinfo {author} {\bibfnamefont
  {Y.}~\bibnamefont {Kuno}}, \bibinfo {author} {\bibfnamefont {H.-T.}\
  \bibnamefont {Wei}}, \bibinfo {author} {\bibfnamefont {R.~T.}\ \bibnamefont
  {Scalettar}}, \bibinfo {author} {\bibfnamefont {K.~R.~A.}\ \bibnamefont
  {Hazzard}},\ and\ \bibinfo {author} {\bibfnamefont {Y.}~\bibnamefont
  {Takahashi}},\ }\href@noop {} {\bibfield  {journal} {\bibinfo  {journal}
  {ArXiv201007730 Cond-Mat}\ } (\bibinfo {year} {2020})},\ \Eprint
  {https://arxiv.org/abs/2010.07730} {arXiv:2010.07730 [cond-mat]} \BibitemShut
  {NoStop}%
\bibitem [{\citenamefont {Fishman}\ \emph {et~al.}(2020)\citenamefont
  {Fishman}, \citenamefont {White},\ and\ \citenamefont
  {Stoudenmire}}]{fishman_itensor_2020}%
  \BibitemOpen
  \bibfield  {author} {\bibinfo {author} {\bibfnamefont {M.}~\bibnamefont
  {Fishman}}, \bibinfo {author} {\bibfnamefont {S.~R.}\ \bibnamefont {White}},\
  and\ \bibinfo {author} {\bibfnamefont {E.~M.}\ \bibnamefont {Stoudenmire}},\
  }\href@noop {} {\bibfield  {journal} {\bibinfo  {journal} {ArXiv200714822
  Cond-Mat Physicsphysics}\ } (\bibinfo {year} {2020})},\ \Eprint
  {https://arxiv.org/abs/2007.14822} {arXiv:2007.14822 [cond-mat,
  physics:physics]} \BibitemShut {NoStop}%
\end{thebibliography}
\end{document}